\documentclass[onecolumn]{aastex62}

\usepackage{times}

\usepackage{amsmath,amssymb,amstext}
\usepackage{tabularx}
\usepackage{longtable}
\usepackage{float}
\usepackage{natbib}

\newcommand{\kms}{km\,s$^{-1}$}

\begin{document}
\title{\large {\bf Multiwavelength Study of Equatorial Coronal-Hole Jets}}
\author{Pankaj Kumar}
\affiliation{Heliophysics Science Division, NASA Goddard Space Flight Center, Greenbelt, MD, 20771, USA}

\author{Judith T. Karpen}
\affiliation{Heliophysics Science Division, NASA Goddard Space Flight Center, Greenbelt, MD, 20771, USA}

\author{Spiro K. Antiochos}
\affiliation{Heliophysics Science Division, NASA Goddard Space Flight Center, Greenbelt, MD, 20771, USA}

\author{Peter F. Wyper}
\affiliation{Department of Mathematical Sciences, Durham University, Durham DH1 3LE, UK}

\author{C. Richard DeVore}
\affiliation{Heliophysics Science Division, NASA Goddard Space Flight Center, Greenbelt, MD, 20771, USA}

\author{Craig E. DeForest}
\affiliation{Southwest Research Institute, 1050 Walnut Street, Boulder, CO, USA}

\email{pankaj.kumar@nasa.gov}

\begin{abstract}
Jets (transient/collimated plasma ejections) occur frequently throughout the solar corona and contribute mass/energy to the corona and solar wind. By combining numerical simulations and high-resolution observations, we have made substantial progress recently on determining the energy buildup and release processes in these jets. Here we describe a study of 27 equatorial coronal-hole jets using Solar Dynamics Observatory/AIA and HMI observations on 2013 June 27-28 and 2014 January 8-10. Out of 27 jets, 18 (67\%) are associated with mini-filament ejections; the other 9 (33\%) do not show mini-filament eruptions but do exhibit mini-flare arcades and other eruptive signatures. This indicates that every jet in our sample involved a filament-channel eruption. From the complete set of events, 6 jets (22\%) are apparently associated with tiny flux-cancellation events at the polarity inversion line, and 2 jets (7\%) are associated with sympathetic eruptions of filaments from neighboring bright points. Potential-field extrapolations of the source-region photospheric magnetic fields reveal that all jets originated in the fan-spine topology of an embedded bipole associated with an extreme ultraviolet coronal bright point. Hence, all our jets are in agreement with the breakout model of solar eruptions. We present selected examples and discuss the implications for the jet energy build-up and initiation mechanisms.

\end{abstract}

\section{INTRODUCTION}
Coronal jets are collimated plasma ejections that occur repeatedly everywhere on the Sun (e.g., in coronal holes, quiet corona, and active regions) and may supply a significant amount of mass and energy to the corona and solar wind \citep{raouafi2016}. Most previous studies of coronal-hole (CH) jets addressed only those events occurring in polar holes and derived their evolving properties solely from extreme ultraviolet/soft X-ray (EUV/SXR) images \citep[e.g.,][]{savcheva2007,cirtain2007,nistico2009,nistico2010,raouafi2010}. Because magnetograms near the limb are of poor quality, the underlying magnetic-field properties of most polar CH jets cannot be determined. In contrast, studies of on-disk, equatorial coronal-hole (ECH) jets are rare, but they benefit critically from access to magnetograms .  

There are no theoretical or observational reasons to expect significant physical differences between equatorial and polar CH jets. Therefore, the lessons learned from polar CH jet studies should apply equally to ECH jets, and vice versa. Two important features of CH jets have emerged recently due to the availability of high-resolution, high-cadence, multiwavelength data: most, if not all, of these events appear to be associated with mini-filament eruptions \citep{sterling2015}; and many exhibit helical, untwisting motions  \citep{patsourakos2008,chandrashekhar2014,innes2016}.  Magnetic reconnection is generally agreed to be the energy-release mechanism, but the energy-storage process and the location and timing of reconnection remain actively debated.  Flux emergence has been  proposed as a driver of coronal jets, through reconnection between the preexisting field and the emerging flux systems \citep[e.g.,][]{shibata1994,moreno2013}. However, as we discuss in \S\ref{results}, few if any CH jets appear to be driven directly by this process. Flux cancellation has been invoked more recently as a mechanism for building up and liberating magnetic free energy in jets \citep[e.g.,][]{panesar2018}. However, a direct connection between ongoing cancellation and the initiation of impulsive jets has not been convincingly demonstrated, in our view.  During flux cancellation, we expect simultaneous and comparable decreases in both fluxes (positive/negative), whereas these studies have only measured the evolution of one polarity. Submergence, diffusion, or fragmentation of flux elements can mimic flux cancellation \citep[e.g.,][]{deforest2007,lamb2013}, particularly for the weak field strengths typical of the jet sources. 

Our previous numerical studies of reconnection-driven coronal jets identified a fundamental magnetic-field topology -- the embedded bipole -- as well as a mechanism of energy buildup and explosive release that yields Alfv\'enic, helical outflows consistent with observations \citep{pariat2009,pariat2010,pariat2015,pariat2016,wyper2016a,wyper2016b,karpen2017}.  We also demonstrated that our breakout model \citep{antiochos1999,karpen2012} for large-scale solar eruptions equally explains small-scale jets \citep{wyper2017,wyper2018} and produces mini-filament eruptions, in agreement with observations by \citet{sterling2015}. Recently we discovered an excellent example of an ECH jet with the classic fan-spine magnetic topology \citep{kumar2018}, characterized by a slowly rising EUV-bright sigmoid and mini-filament, dimmings at both ends of the sigmoid, weak quasiperiodic outflows at the null, multiple plasmoid formation in the flare current sheet beneath a rapidly rising flux rope, and jet onset resulting from explosive breakout reconnection between the flux rope and the external open field. There was no evidence of flux emergence or cancellation up to 16 hours before the impulsive event. For this case, the observed features closely matched the predictions of our breakout-jet model. 

To establish whether these results are generally applicable to CH jets, we identified 27 well-observed on-disk jets in two equatorial coronal holes and analyzed their EUV coronal emissions and photospheric magnetic-field evolution. In this paper, we report the results of this study, which closely agree with the predictions of the breakout-jet model and do not support the flux-emergence or -cancellation scenarios for explosive energy release. After describing the data selection and analysis methods (\S \ref{obs}), we present observations of the evolving jet source regions and selected examples of ECH jets with and without filament eruptions (\S \ref{results}). In \S \ref{conclusions}, we summarize our conclusions regarding the pre-event configuration, roles of flux emergence and cancellation, and evidence for the breakout model.

\section{DATA SELECTION AND ANALYSIS}\label{obs}

We used the {\it Solar Dynamics Observatory} (SDO)/Atmospheric Imaging Assembly (AIA; \citealt{lemen2012}) full-disk images of the Sun (field-of-view $\approx$1.3~R$_\odot$) with a spatial resolution of 1.5$\arcsec$ (0.6$\arcsec$~pixel$^{-1}$) and a cadence of 12~s, in the following channels: 304~\AA\ (\ion{He}{2}, at temperature $T\approx 0.05$~MK), 171~\AA\ (\ion{Fe}{9}, $T\approx 0.7$~MK), 193~\AA\ (\ion{Fe}{12}, \ion{Fe}{24}, $T\approx  1.2$~MK and $\approx 20$~MK), and 211~\AA\ (\ion{Fe}{14}, $T\approx 2.0$~MK) images. We also analyzed cotemporal SDO/Helioseismic and Magnetic Imager (HMI; \citealt{schou2012}) magnetograms at a 45-s cadence. A new 3D noise-gating technique \citep{deforest2017} was used to clean the AIA images and the HMI magnetograms.

\begin{figure*}
\centering{
\includegraphics[width=8.2cm]{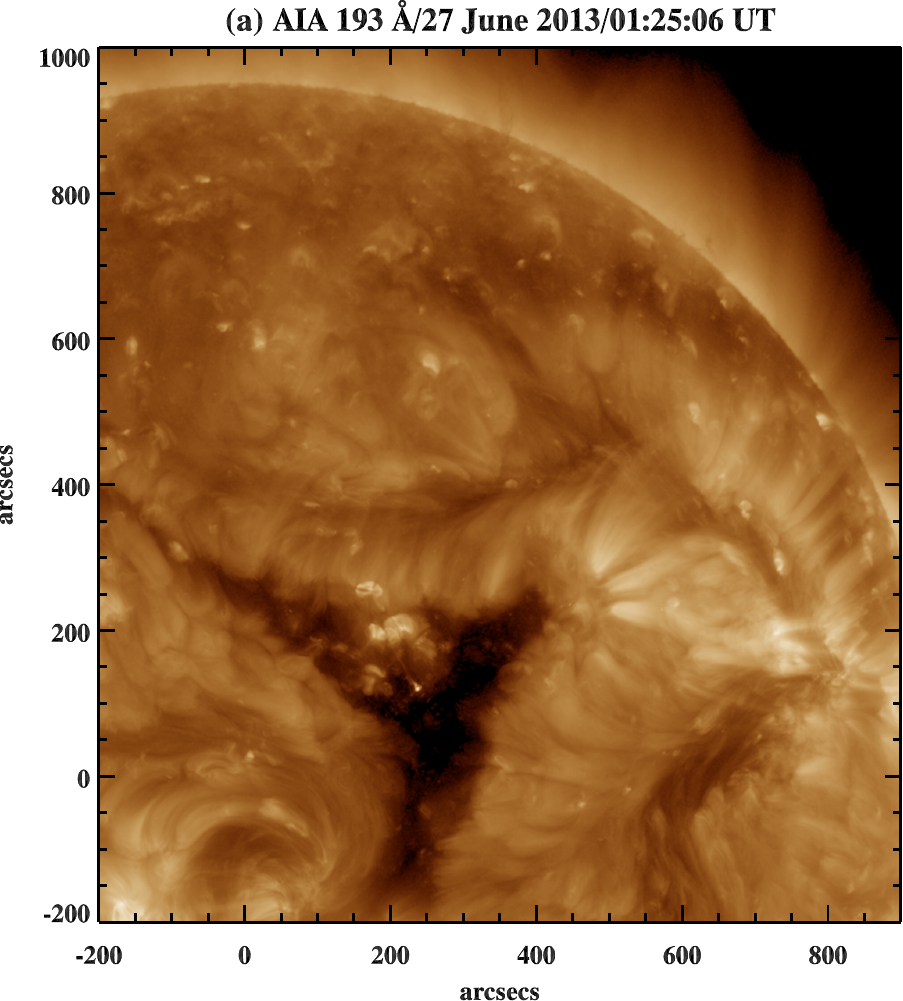}
\includegraphics[width=7.5cm]{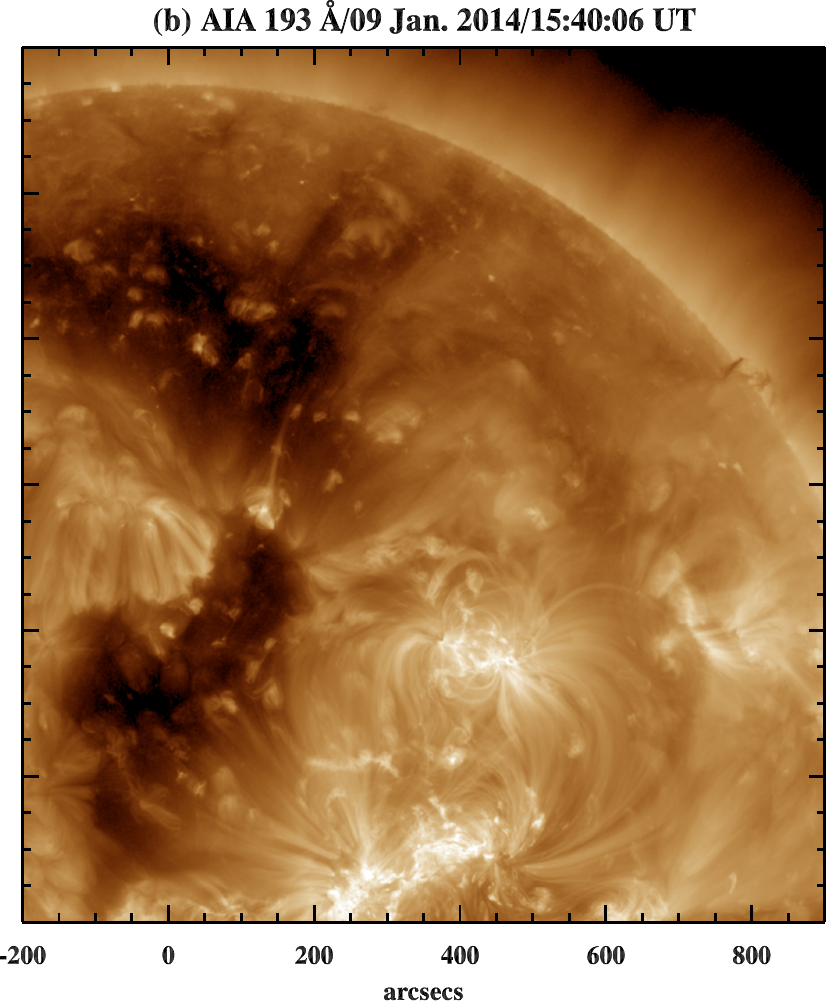}
}
\caption{Views from AIA's 193~\AA~ channel of two equatorial coronal holes from which our set of ECH jets was identified and analyzed. (a) 2013 June 27. (b) 2014 January 9.} 
\label{ech}
\end{figure*}

We selected 27 jets from large, well-observed ECHs on 2013 June 27-28 and 2014 January 8-10 (see Figure \ref{ech}). After viewing SDO/AIA movies of the ECHs, we mainly selected bigger jets so that we could study their magnetic-field topology, evolution of the photospheric magnetic field, and coronal structures before, during, and after eruption. The AIA 304, 171, and 193 \AA~ movies also showed key features such as mini-filaments, jet onset, and flare ribbon/arcade formation. For each event, Table 1 lists the jet's number, date, eruption start and end times, brief description, onset time for the mini-flare arcade, whether or not a mini-filament existed in the source region, whether or not flux cancellation was observed within 3 hours of jet onset, time delay between initial bright-point emergence and the first jet onset, and links to AIA movies. Within the regions of interest in the HMI magnetograms, we measured the evolving positive and negative photospheric fluxes above a threshold of $\pm$30 G for $\approx$3 hours before the jet onset. In our experience this method provides more robust estimates of local magnetic flux changes in both polarities than simply following one polarity with time, particularly if the goal is to determine whether flux cancellation plays an important role in generating jets. Potential-field extrapolations from pre-event HMI magnetograms were used to estimate the magnetic structure of the jet sources and surroundings. To reveal dimming regions associated with the selected jets, we created movies of AIA 193 \AA~ base-difference images. To determine the kinematics of the rising structures and outflows, we extracted EUV intensity profiles from the AIA 193 \AA~ images along narrow slits placed on the paths of the rising structures and created time-distance (TD) intensity plots. AIA 193 \AA~ mean counts were extracted from selected portions of the TD plots to represent the temporal evolution of the flare emission associated with our jets; the measured counts were averaged over the selected portions in order to increase the signal-to-noise ratio. Similarly, we created TD flux plots of magnetic field strength within a confined region around the PIL in each embedded bipole to detect changes in the relative locations of positive and negative polarity concentrations, as an indicator of flux cancellation. 

\begin{figure*}
\centering{
\includegraphics[width=5cm]{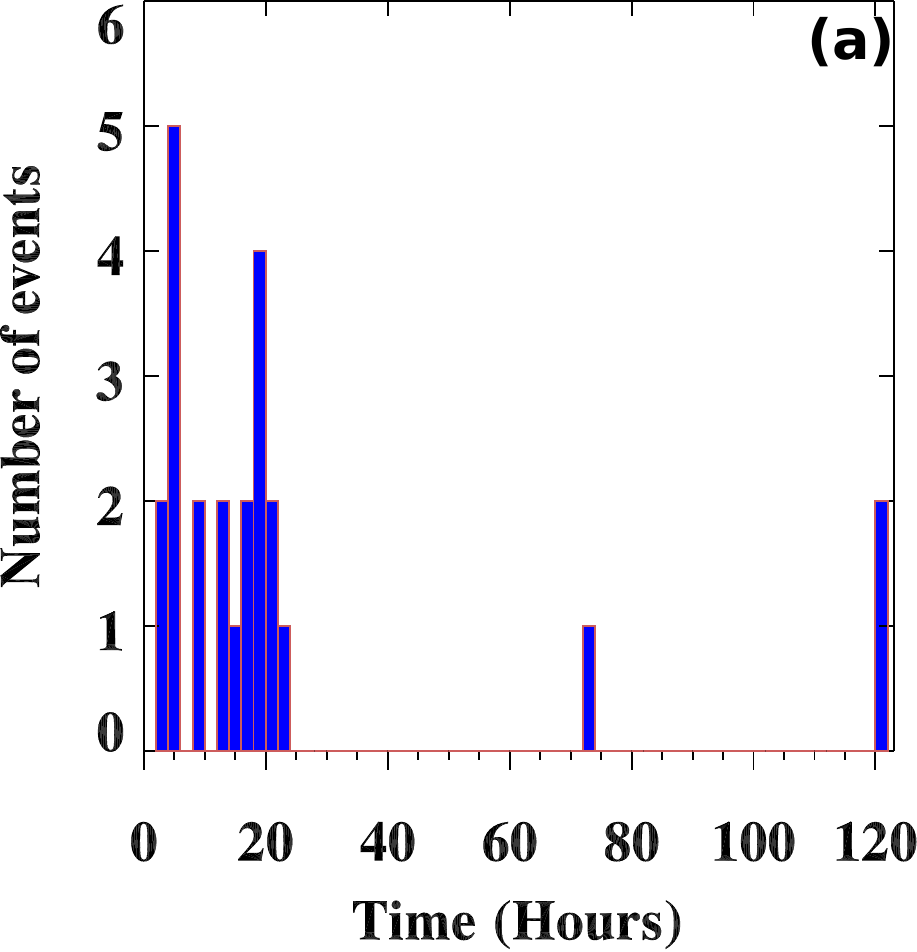}
\includegraphics[width=5cm]{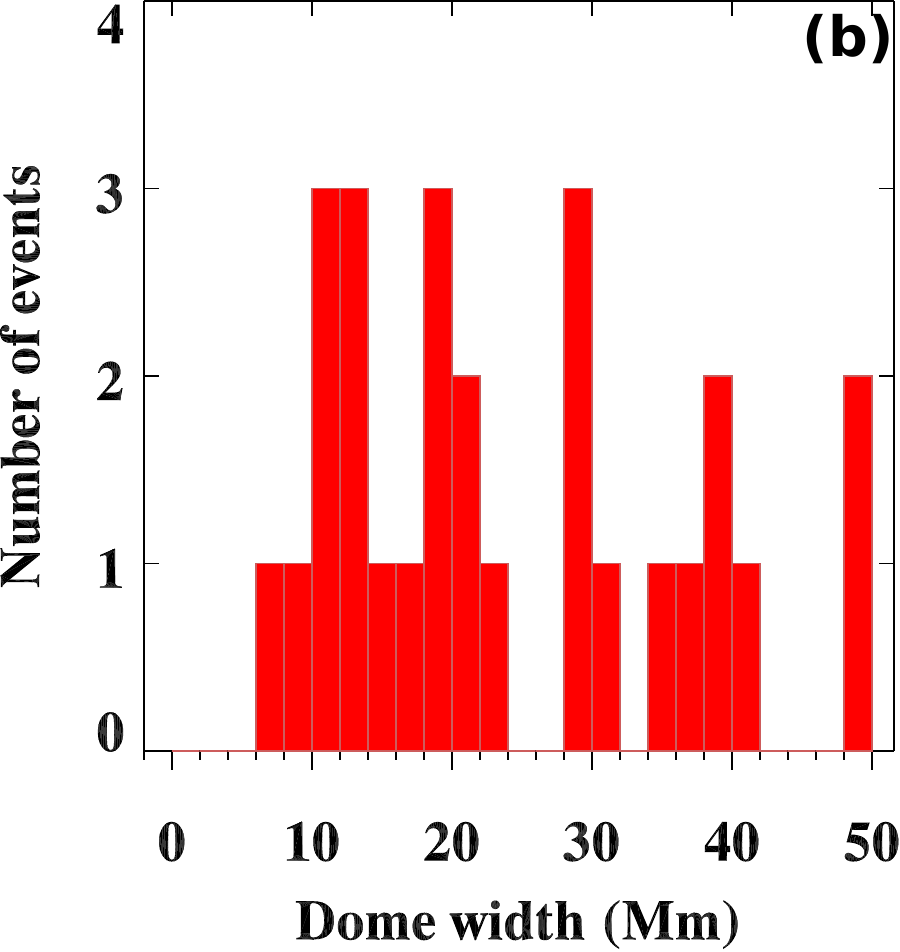}
\includegraphics[width=5cm]{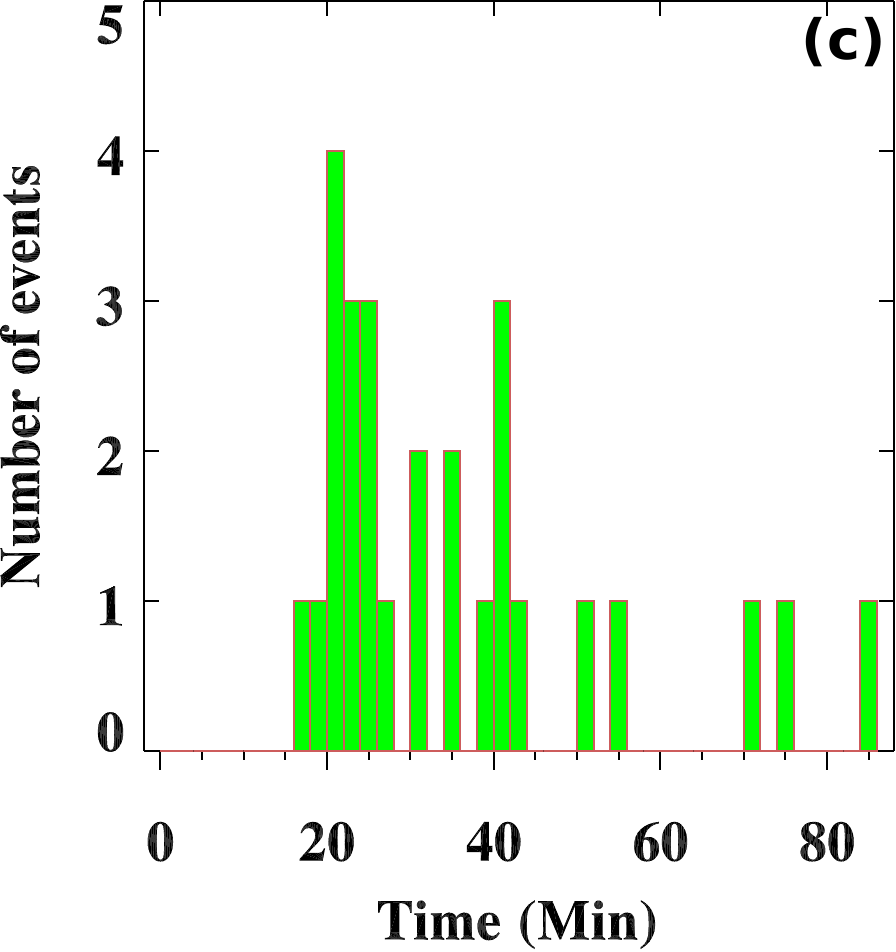}
\
}
\caption{Frequency distributions of the equatorial coronal hole jets properties. (a) Time interval between bright-point appearance and first jet. (b) Dome width measured in the AIA 193 \AA~ channel. (c) Jets duration estimated from the eruption onset to the disappearance of jets spire in the AIA 193 \AA.   
} 
\label{stat}
\end{figure*}
\begin{figure*}
\centering{
\includegraphics[width=5.75cm]{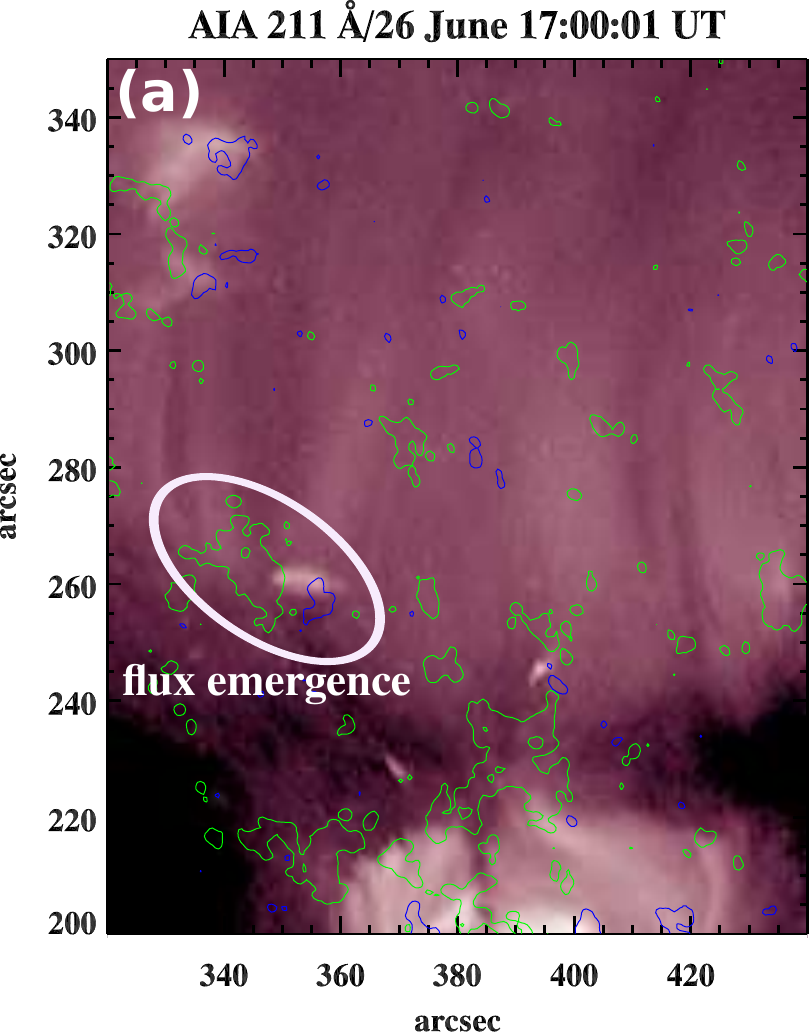}
\includegraphics[width=5cm]{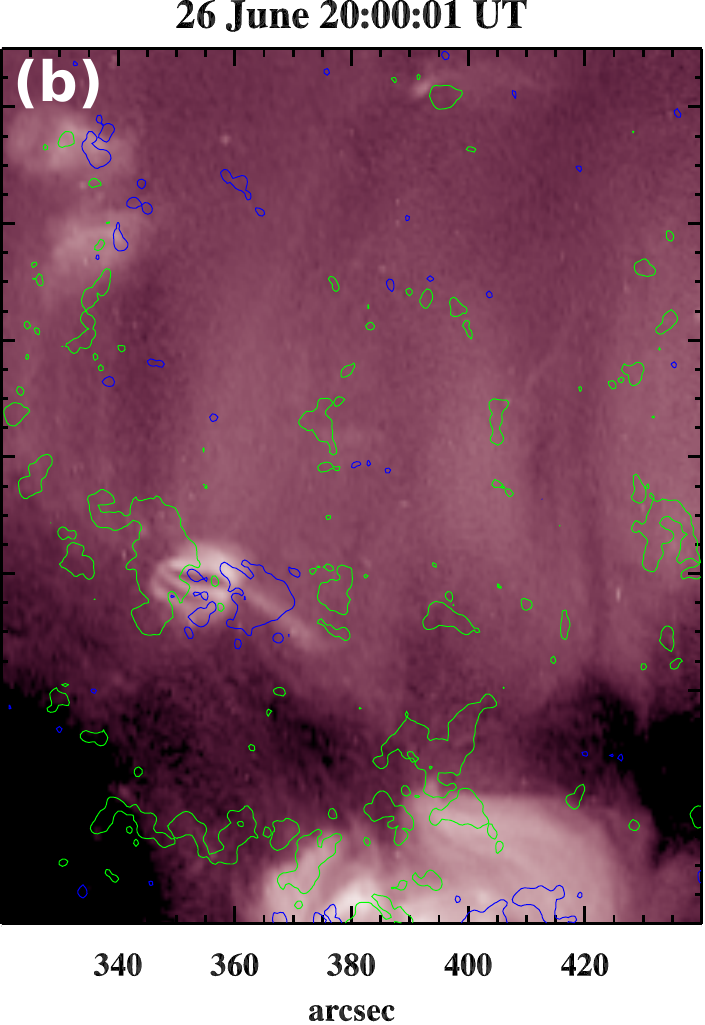}
\includegraphics[width=5cm]{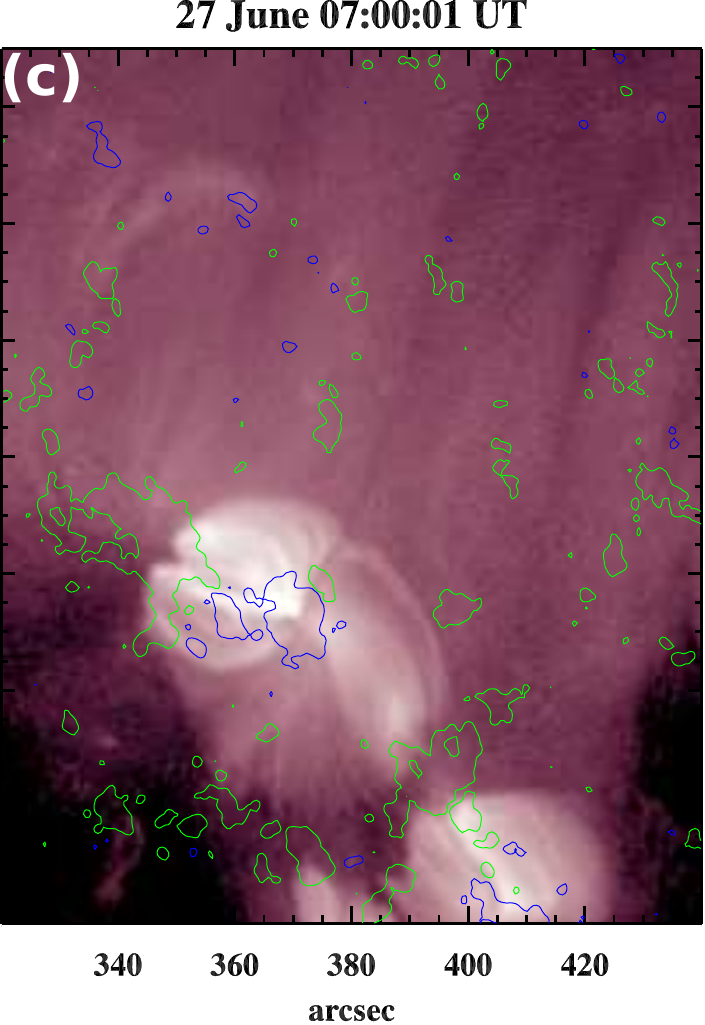}
\includegraphics[width=5.75cm]{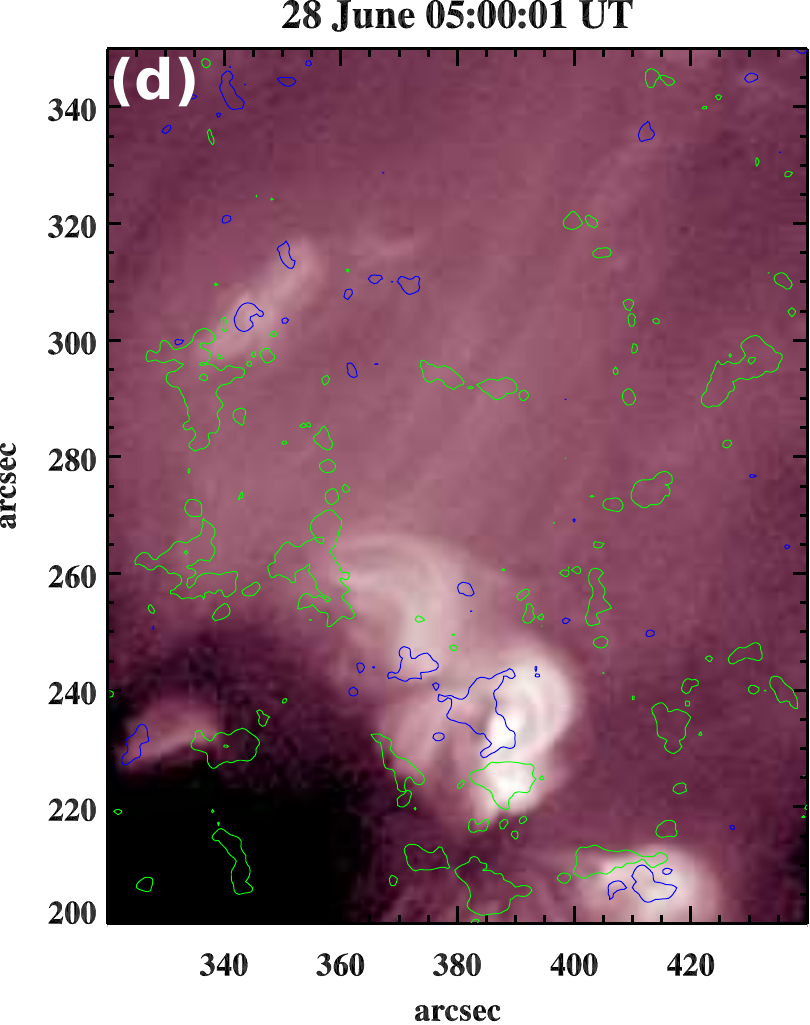}
\includegraphics[width=5cm]{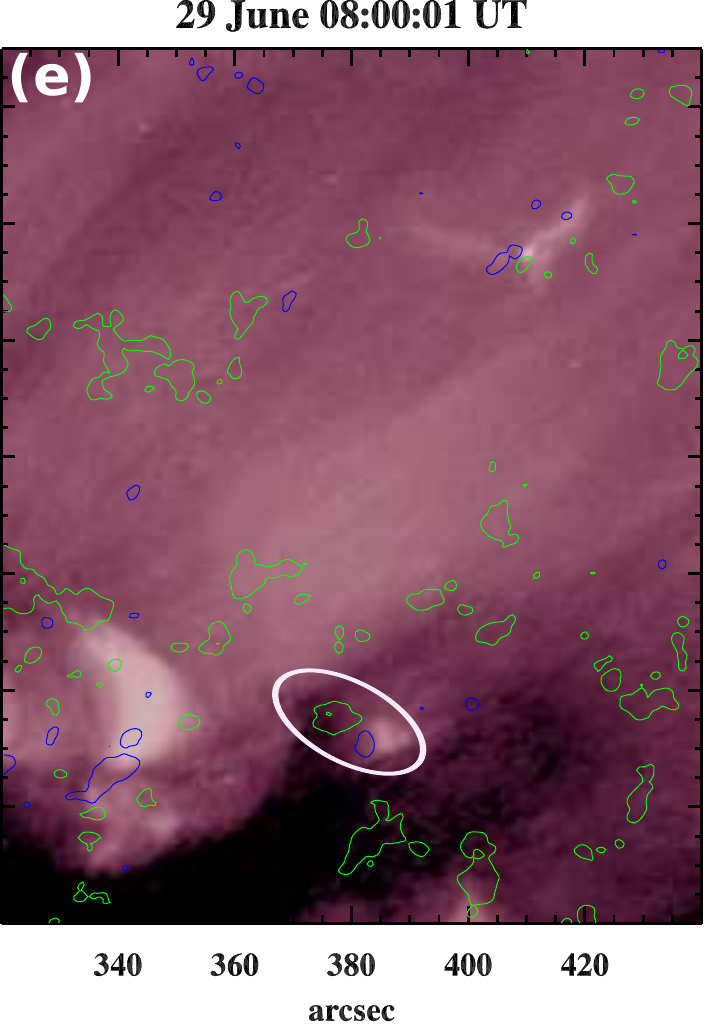}
\includegraphics[width=5.22cm]{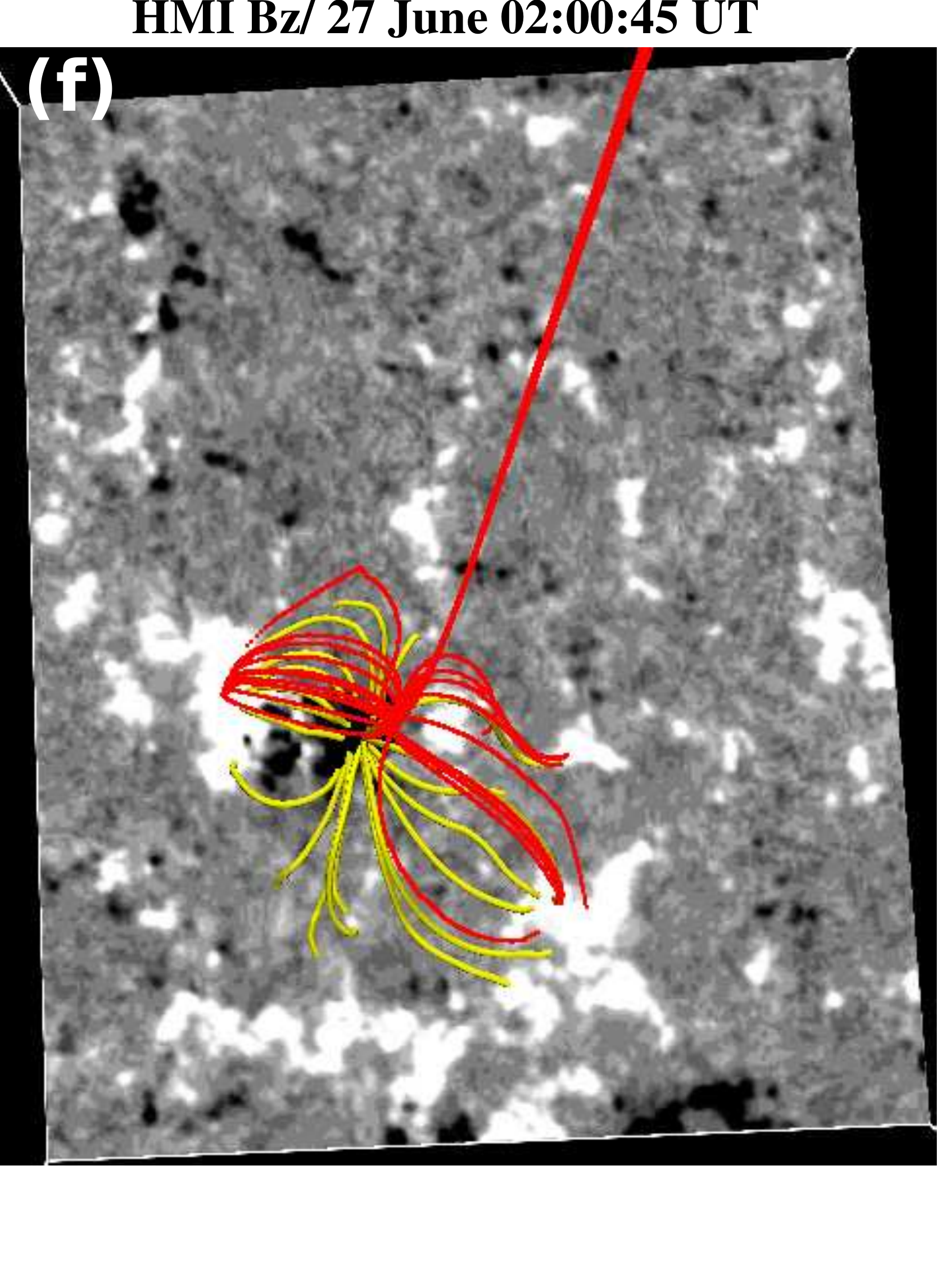}
}
\caption{An example of the evolution of an ECH bright point from its appearance (2013 June 26) to disappearance (2013 June 29). This region produced many diffuse jets, which are not in Table 1. (a-e) The AIA 211 \AA~ images at selected times are overlaid by the cotemporal HMI magnetogram contours with $\pm$30 G (green=positive, blue=negative).  (f) HMI magnetogram prior to onset of diffuse jets. Selected closed (yellow) and open (red) field lines from a Potential-field extrapolation delineate the fan and spine of the underlying embedded bipole. (An animation of this figure is available).  
} 
\label{bp}
\end{figure*}


\begin{figure*}
\centering{
\includegraphics[width=18cm]{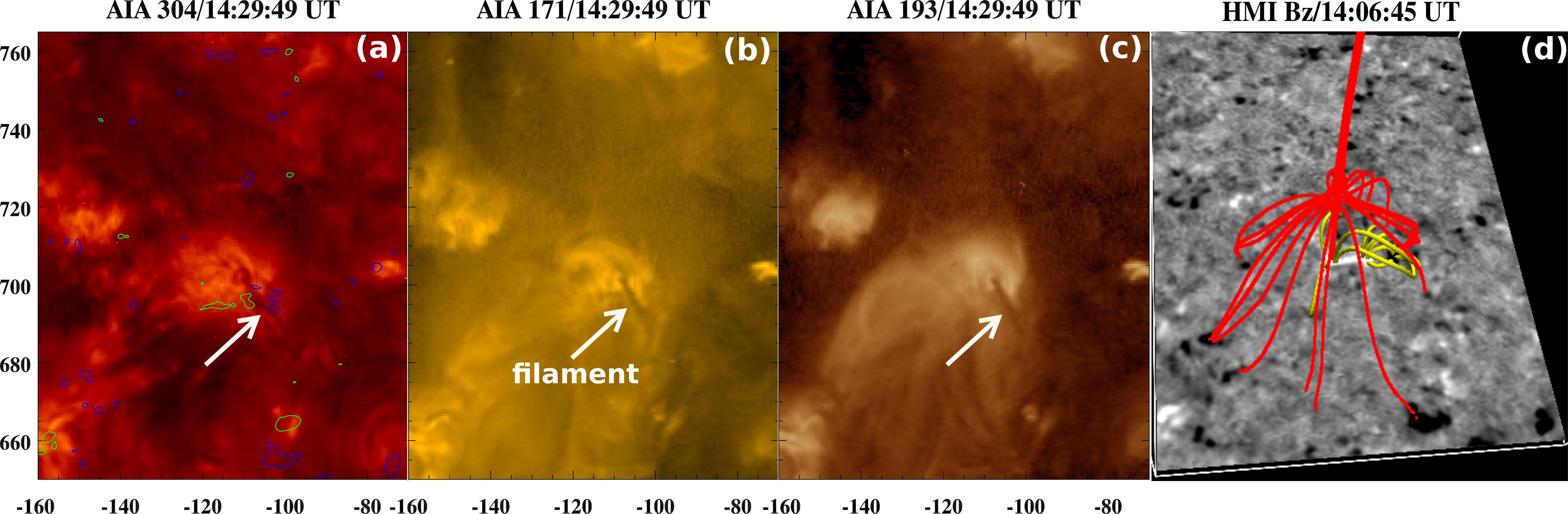}
\includegraphics[width=4.55cm]{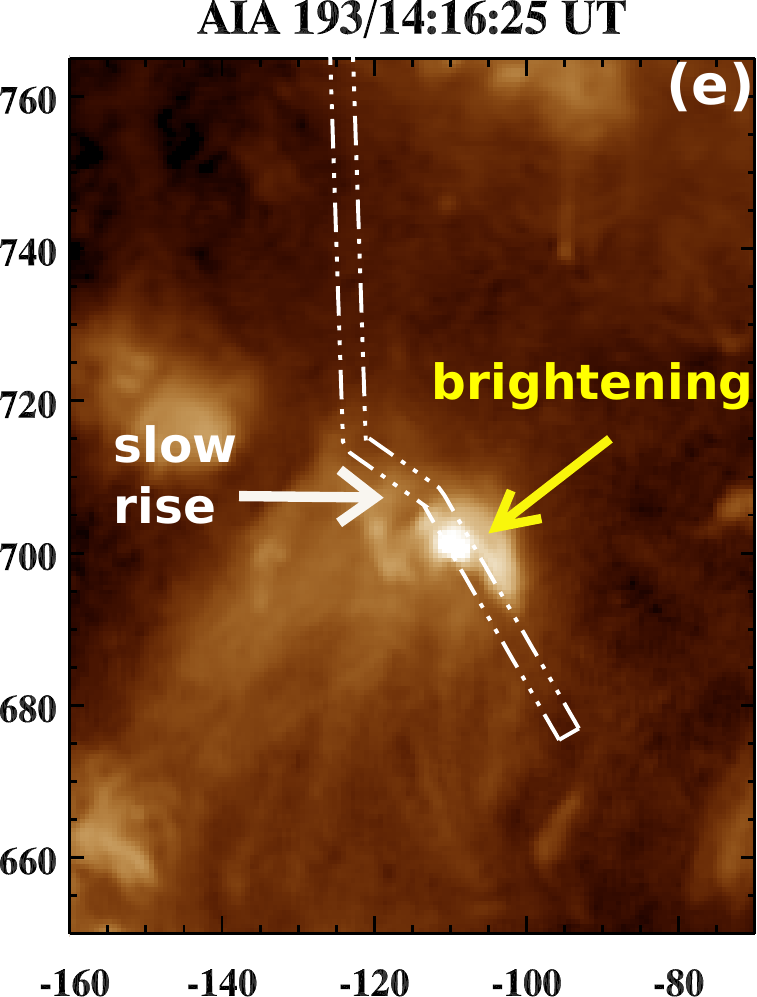}
\includegraphics[width=4.3cm]{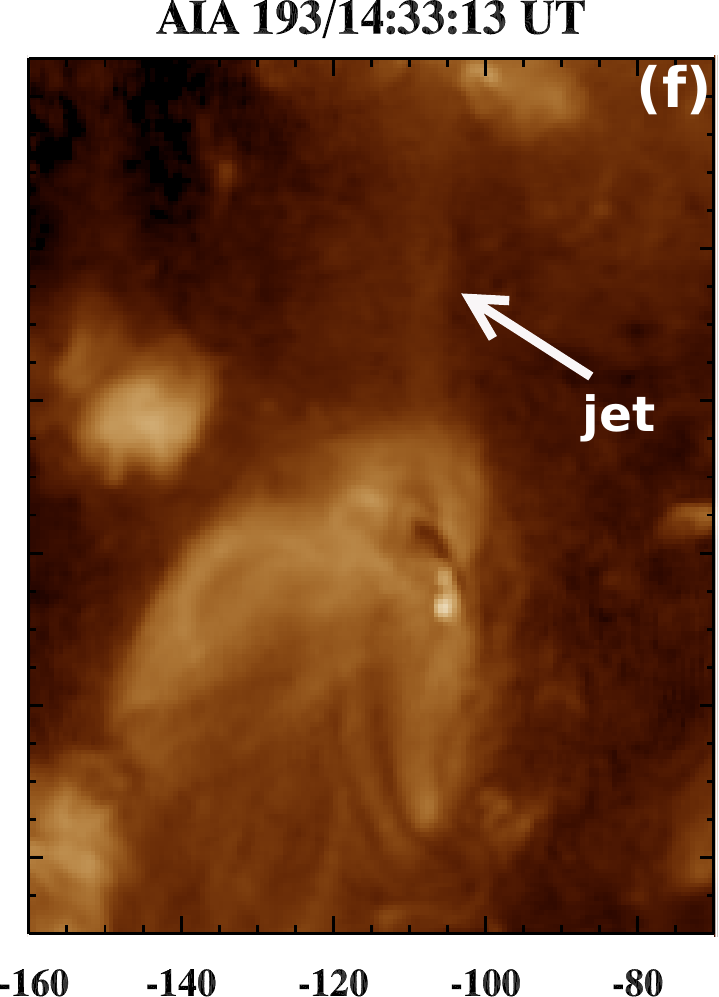}
\includegraphics[width=4.3cm]{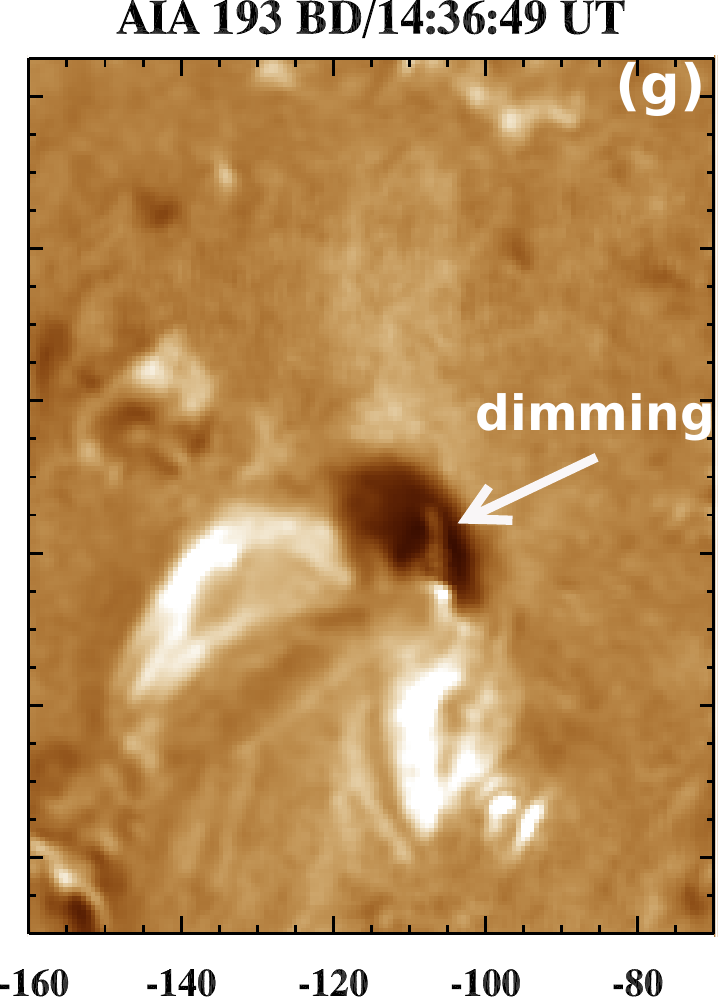}
\includegraphics[width=4.3cm]{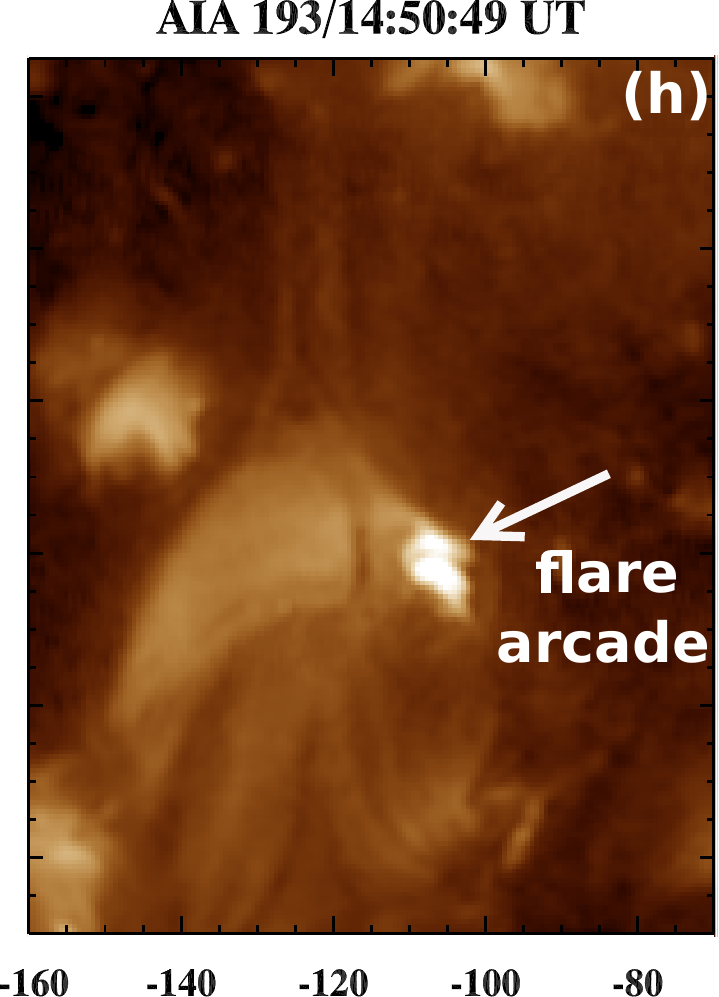}
\includegraphics[width=19cm]{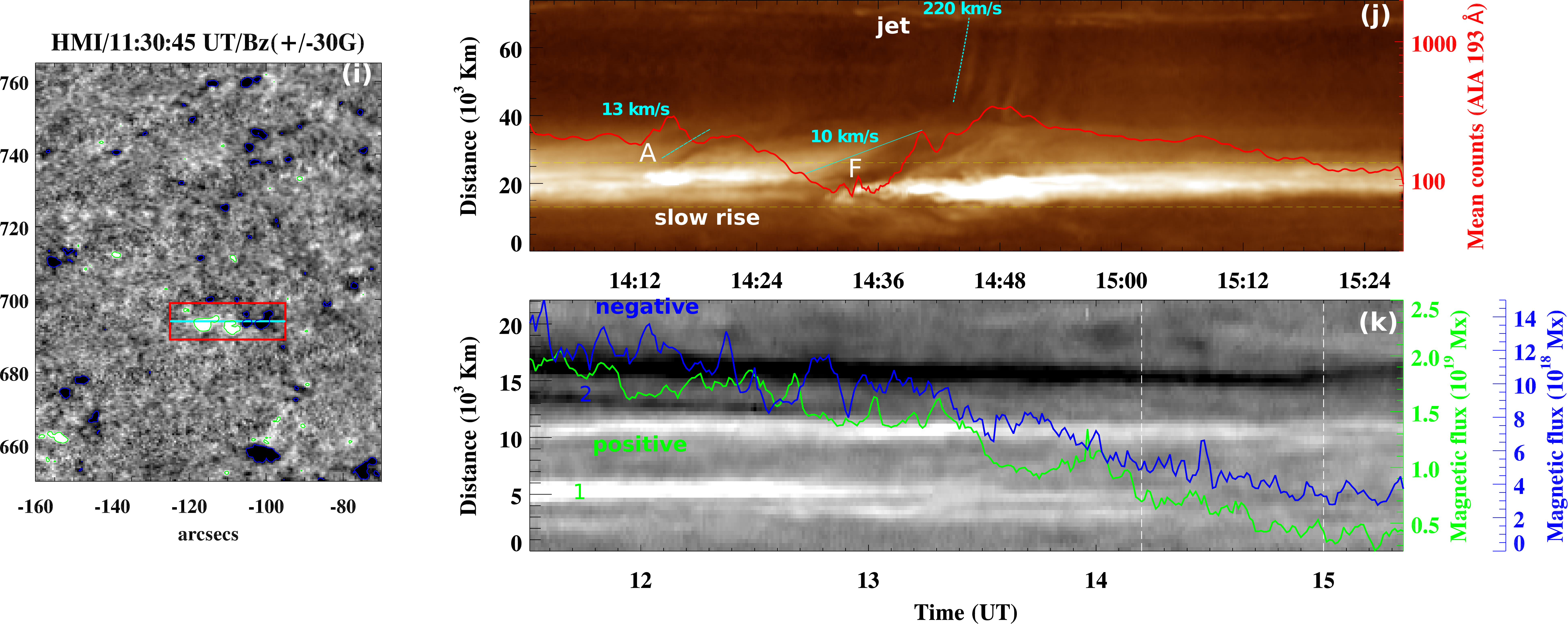}
}
\caption{(a-c) AIA 304, 171, and 193 \AA~ images of a jet associated with a mini-filament eruption (\#7 in Table 1).  HMI magnetogram contours ($\pm$30 G) of positive (blue) and negative (green) polarities are superposed on the EUV image.  (d) Potential-field extrapolation of the jet source region. (e-h) AIA 193 \AA~ images showing the slow rise, onset of a diffuse jet, formation of a dimming region, and the mini-flare arcade. (g) Base difference image (14:36:49 UT-14:01:37 UT). (i) HMI magnetogram with $\pm$30 G contours prior to the jet onset. (j) TD intensity plot extracted from AIA 193 \AA~ images along the slit (white dot-dashed line) shown in panel (e). The red curve represents the averaged counts extracted between the two horizontal dashed lines (yellow). (k) TD flux plot from a 4-hr series of HMI magnetograms, taken along the cyan cut shown in (i) and averaged in the north-south direction within the red box. Blue and green curves are the negative and positive fluxes (absolute value) in Mx within the $\pm$30 G contours inside the red box in (i). Vertical white dashed lines mark the beginning and end of the eruption phase (An animation of this figure is available).   
} 
\label{j1}
\end{figure*}


\section{RESULTS}\label{results}

\subsection{Energy buildup: appearance and disappearance of coronal bright points}

\begin{figure*}
\centering{
\includegraphics[width=18cm]{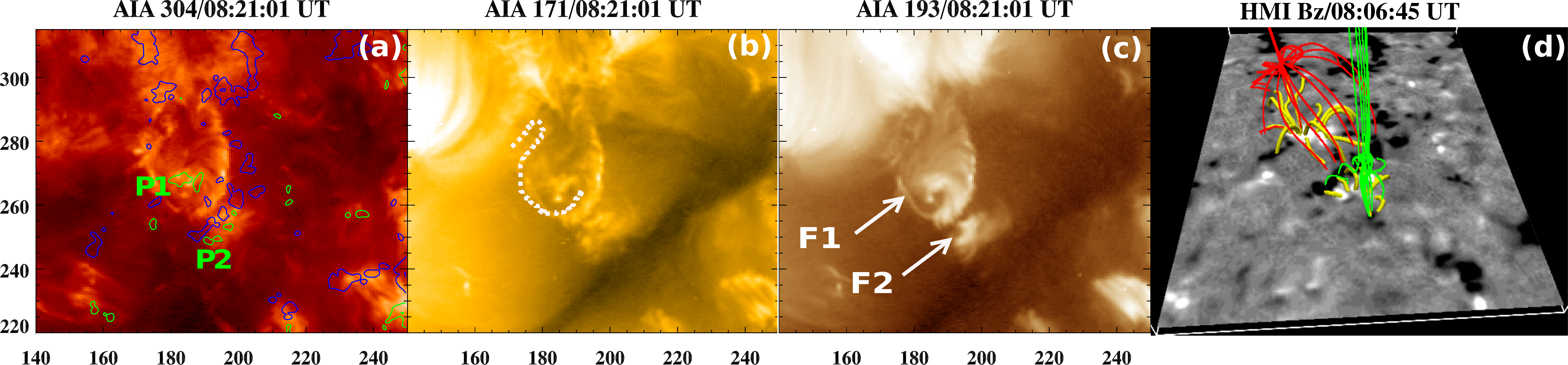}
\includegraphics[width=4.65cm]{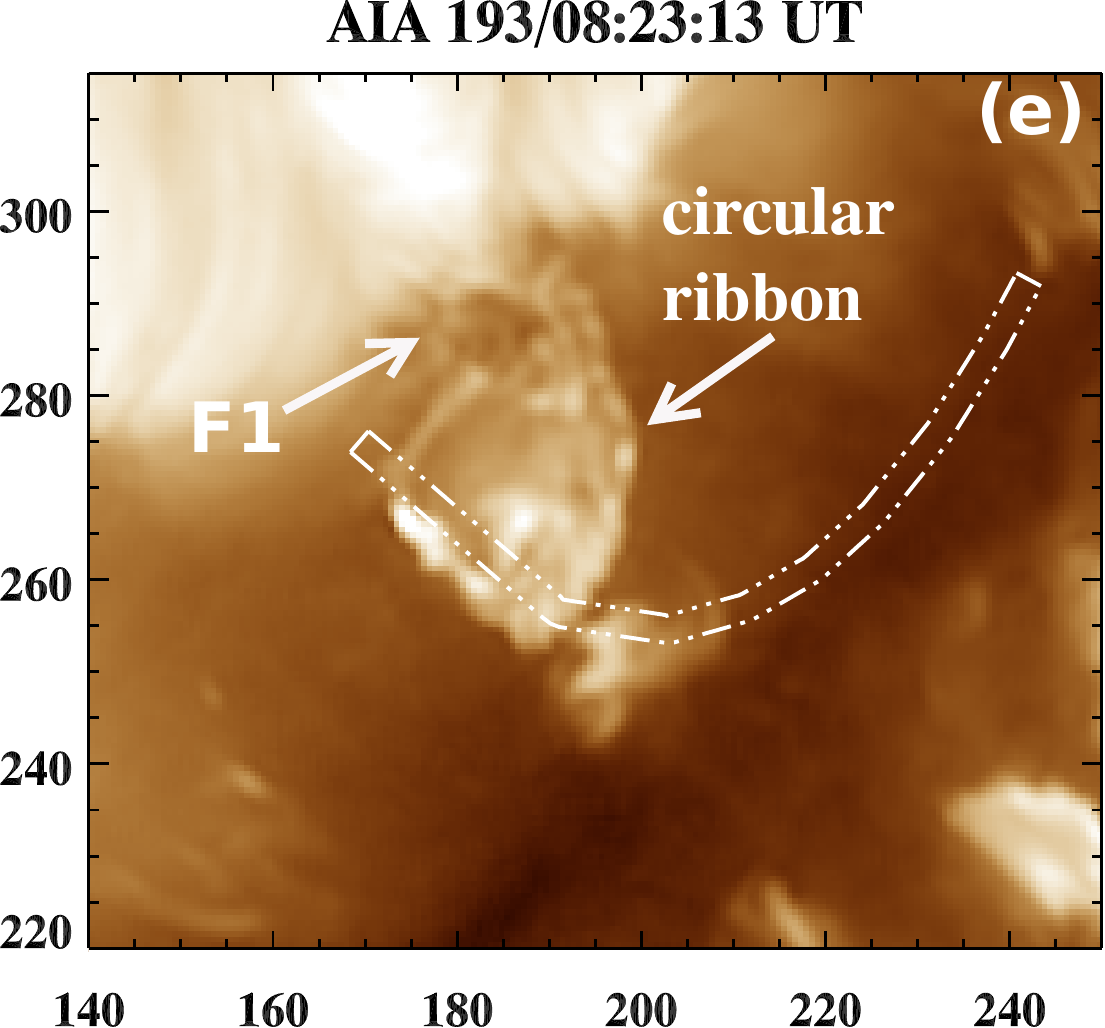}
\includegraphics[width=4.3cm]{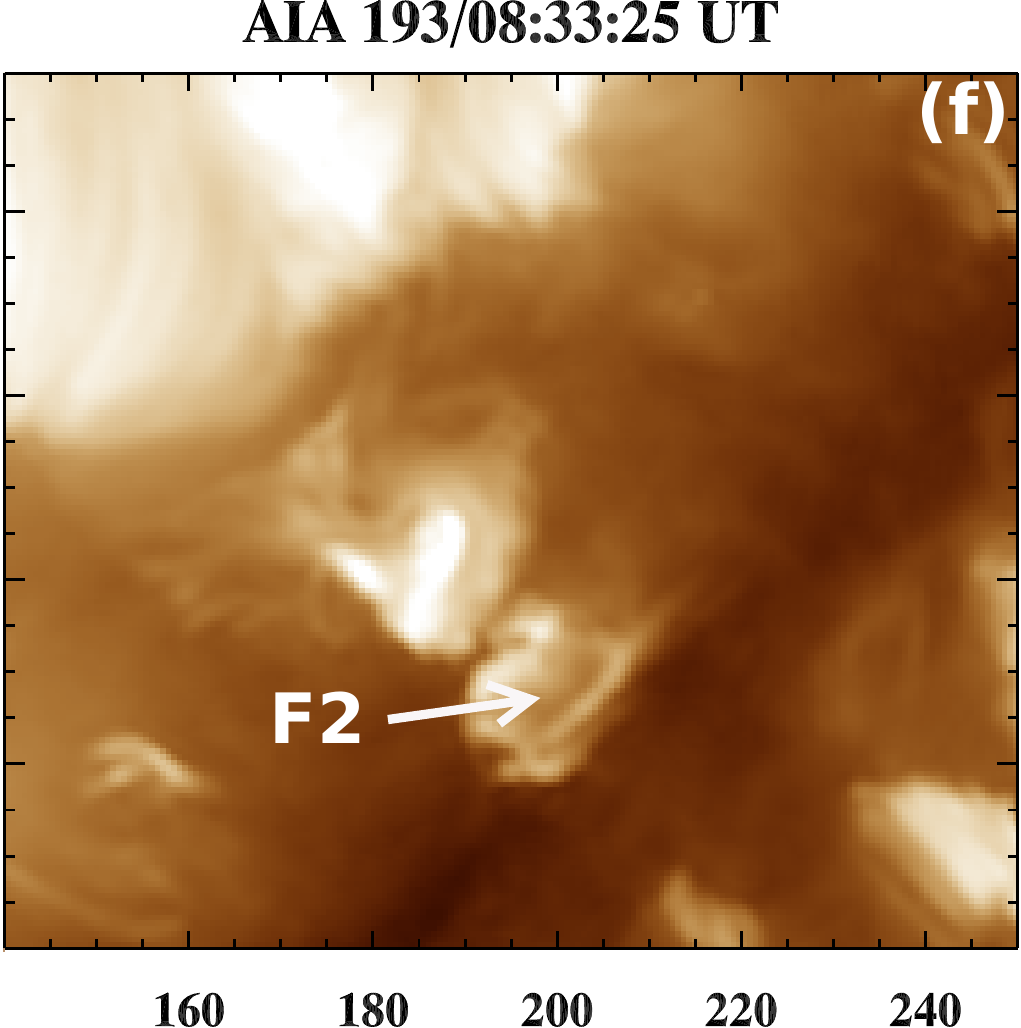}
\includegraphics[width=4.3cm]{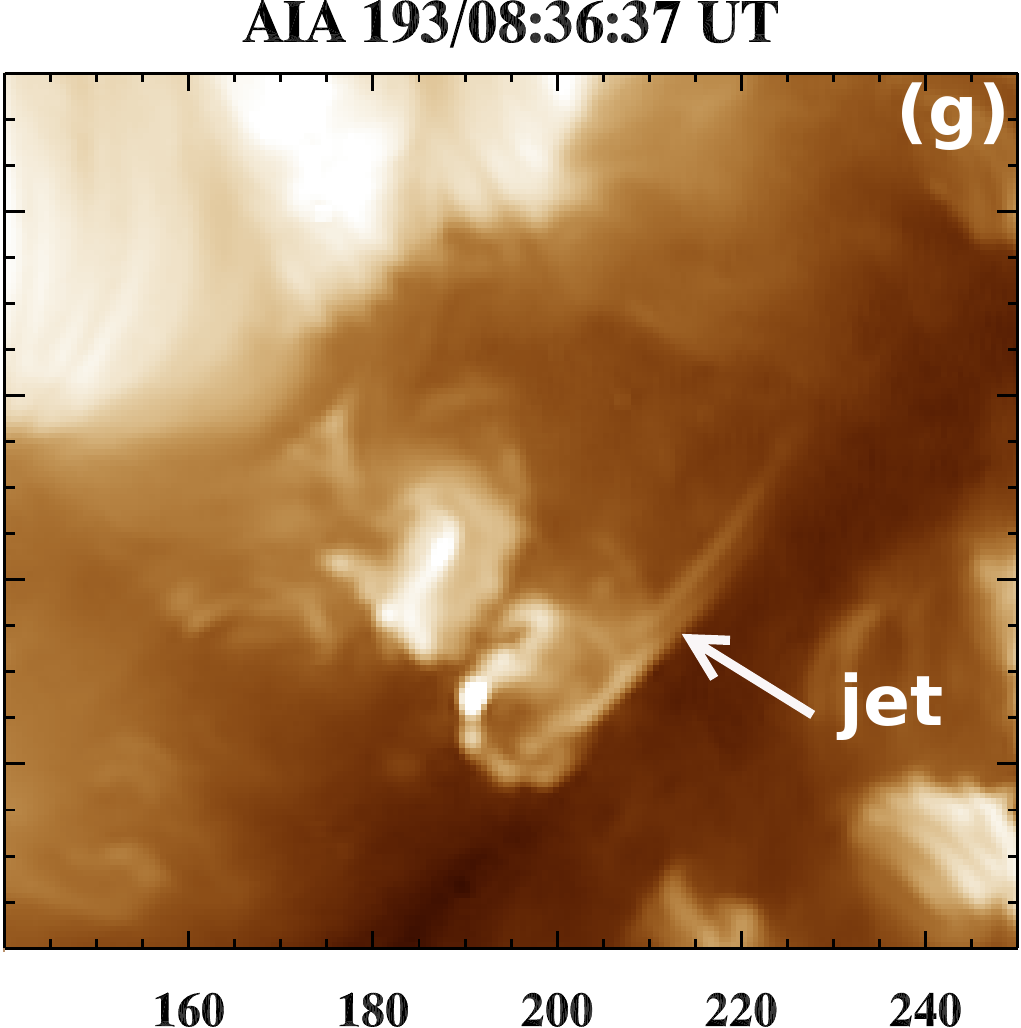}
\includegraphics[width=4.3cm]{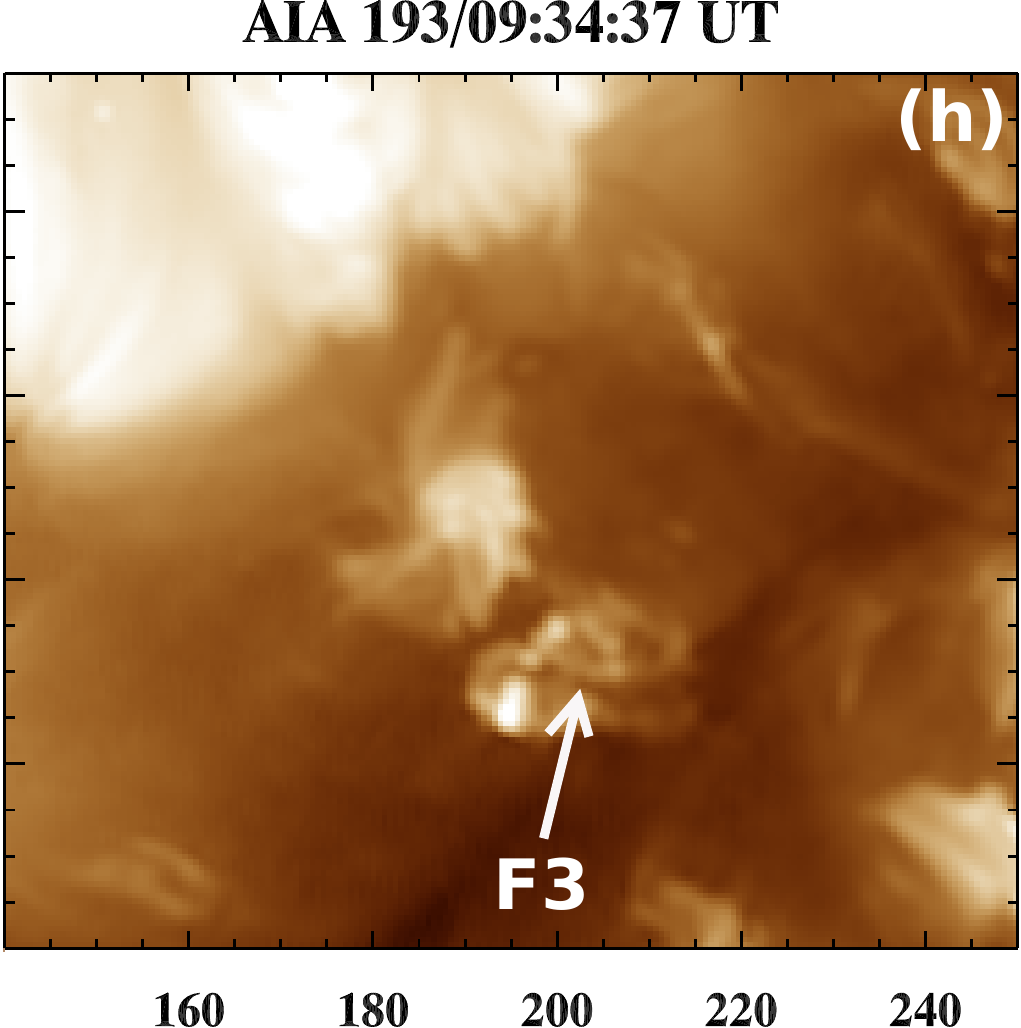}
\includegraphics[width=19cm]{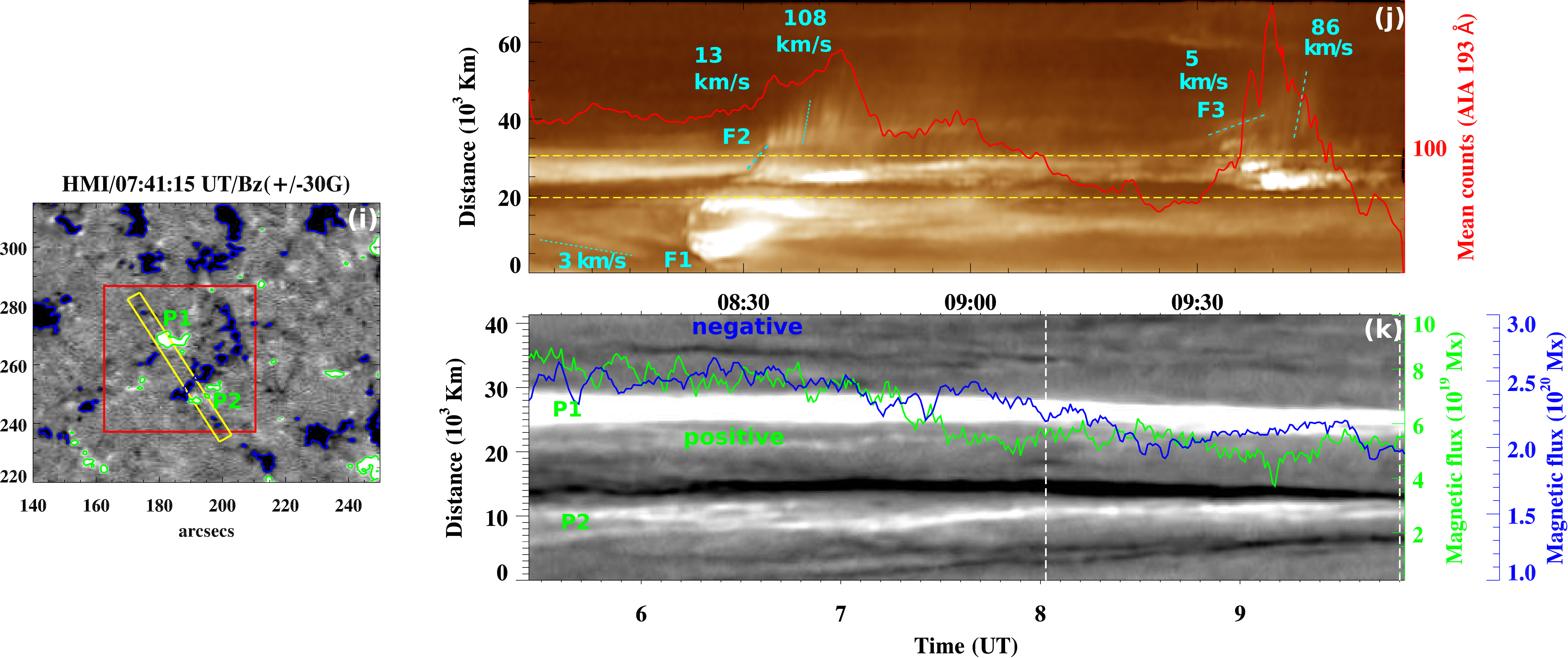}
}
\caption{(a-c) AIA 304, 171, and 193 \AA~ images of two neighboring bright points that generated 3 jets (\#17-19 in Table 1) associated with the eruption of mini-filaments F1, F2, and F3.  HMI magnetogram contours ($\pm$30 G) of positive (blue) and negative (green) polarities are superposed on on panel (a). P1 and P2 are the minority-polarity regions (positive) surrounded by CH background (negative) magnetic field. (d) Potential-field extrapolation of the jet sources, showing selected closed (yellow) and open (red and green) field lines.  (e-h) AIA 193 \AA~ images showing the eruption of filaments F1, F2, and F3 associated with Jets \#17, 18 and 19, respectively. (i) HMI magnetogram with $\pm$30 G contours prior to the first jet onset. (j) TD intensity plot extracted from AIA 193 \AA~ images along the slit (white dot-dashed line) shown in panel (e). The red curve represents the averaged counts taken from between the two horizontal dashed lines (yellow). Note that the curved slit and the southward direction of the F1 jet yield a downward-directed feature for the F1 eruption in this TD plot, which should not be interpreted as a downflow. (k) TD flux plot from a 5-hr series of HMI magnetograms, taken from the narrow yellow box in panel (i) and averaged along its width.  Blue and green curves are the negative and positive fluxes (absolute value) within the $\pm$30 G contours inside the red box in panel (i). Vertical white dashed lines mark the beginning and end of the eruption phase (An animation of this figure is available).   
} 
\label{twin}
\end{figure*}

To explore the role of emerging flux in the production of CH jets, we analyzed AIA 193 \AA~  images and HMI magnetograms from the first appearance of the source regions until their disappearance.  These sources are all coronal bright points, which are well-known signatures of flux emergence \citep[e.g.,][]{golub1980} best seen in the AIA 193 and 211 \AA~ channels. Our observations indicate that the appearance and disappearance of CH bright points are associated with the emergence and dispersal (diffusion/cancellation/submergence), respectively, of embedded bipoles, with the underlying magnetic structure traced by bright EUV and/or soft X-ray loops \citep[e.g.,][]{kumar2015a}. The bright points do not produce jets for a significant interval after emerging ($\approx$2 hours for the smallest bright points and $\approx$5 days for the largest, in our sample), suggesting that the buildup of sufficient free energy at the PIL takes hours at a minimum (Figure \ref{stat}(a)).  During this interval the EUV loops evolve from a bipolar to an ``anemone" configuration, reflecting  changes in the connectivity: the minority polarity is initially connected only to its bipolar counterpart, then it establishes connections to the surrounding majority-polarity concentrations to form the classic fan-spine topology of a bright point. In topological terms, the null point rises from the low photosphere to high in the corona. Multiple eruptions may occur until the free energy has been spent and/or the underlying magnetic structure has become too diffuse to erupt.  Frequently, the bright point and associated minority-polarity region disappear soon after the last jet.  We also found that the bright points associated with stronger magnetic fields and more volume persist longer than the more compact bright points with weaker fields. For instance, the northern bright point and central minority polarity that produced jet \#17 were visible longer than their southern counterparts where jets \#18 and \#19 originated (see Figure \ref{twin}). This pattern is consistent with the distribution of active-region lifetimes as a function of size and magnetic-field strength \citep[see review by][]{lidia2015}. The lifetimes of the studied jet sources ranged from 7 hours to at least 6 days (an entire disk passage), while the diameters of the bright points at the times of the primary jets ranged from 7 to 48 Mm (Figure \ref{stat}(b)). The duration of jets (Figure \ref{stat}(c)) is estimated using AIA 193 \AA~ movies; from the eruption onset time (slow rise with internal brightening) to the disappearance of jet spire. Most of the events duration ranges between 20-50 min, which is consistent with the previous studies \citep[e.g.][]{nistico2009,nistico2010}. The magnetic field evolves constantly during the bright-point life cycle. In addition to the usual signs of emergence, the central minority polarity moves translationally, rotates, and breaks up or disappears after emergence has stopped. Encounters with opposite-polarity concentrations often appear to produce slow cancellation, but our study shows that this cancellation is uncorrelated with explosive coronal activity (\S\S3.2 and 3.3)
 
To illustrate these features, Figure \ref{bp} and its accompanying animation show the evolution of a bright point in a $\approx$65-hour movie of the AIA 211 \AA~ images and HMI magnetograms during 2013 June 26-29.  The bright point appeared during the emergence of a small bipole (marked in white oval) within the CH background field (Figure \ref{bp}(a)). By 20:00 UT on June 26, fan loops began to connect the central minority-polarity region P1 (positive) with surrounding opposite-polarity regions (Figure \ref{bp}(b)-(d)). The first jet from this bright point was detected after $\approx$45 hours of emergence. The bright point produced recurrent jets before disappearing (within white oval in Figure \ref{bp}(e)) as the minority polarity dispersed. 

\begin{figure*}
\centering{
\includegraphics[width=18cm]{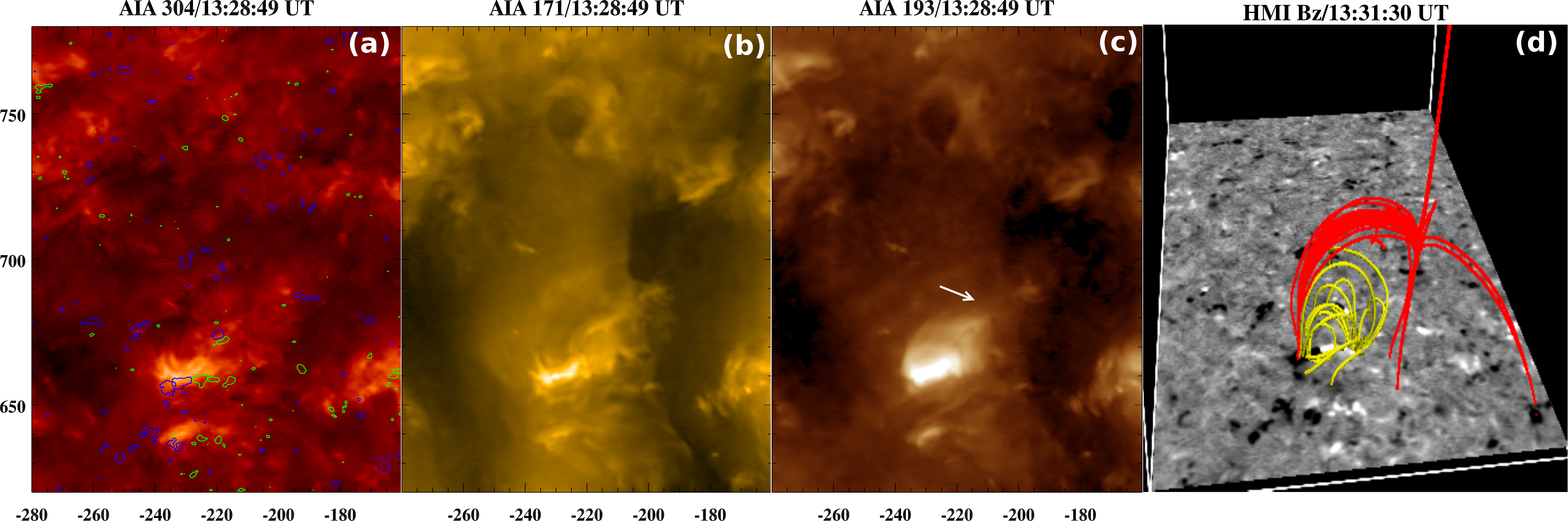}
\includegraphics[width=4.45cm]{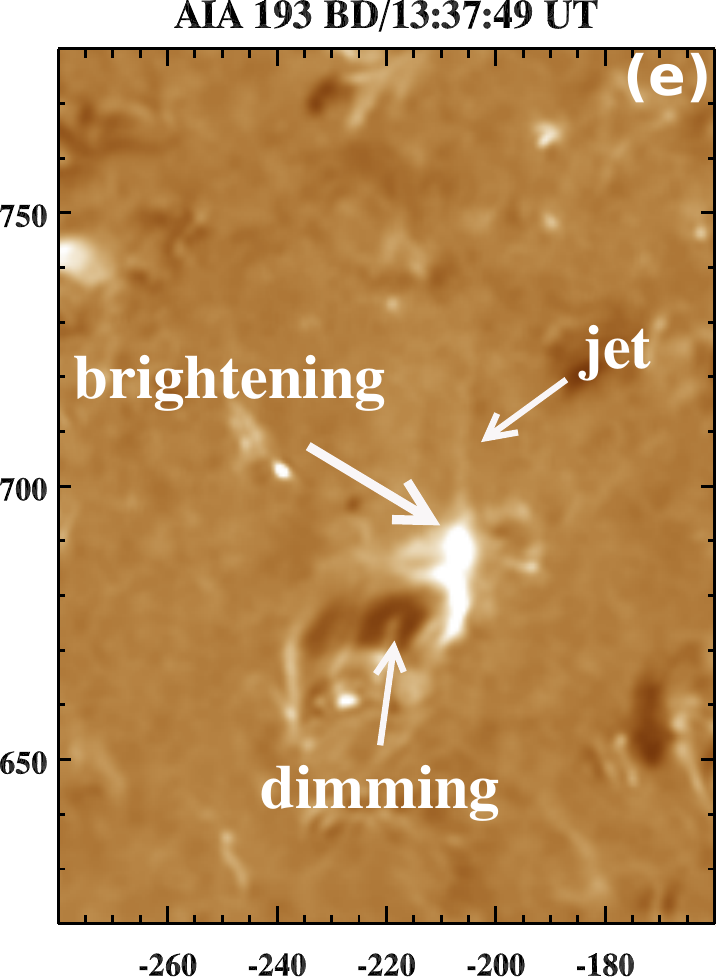}
\includegraphics[width=4.1cm]{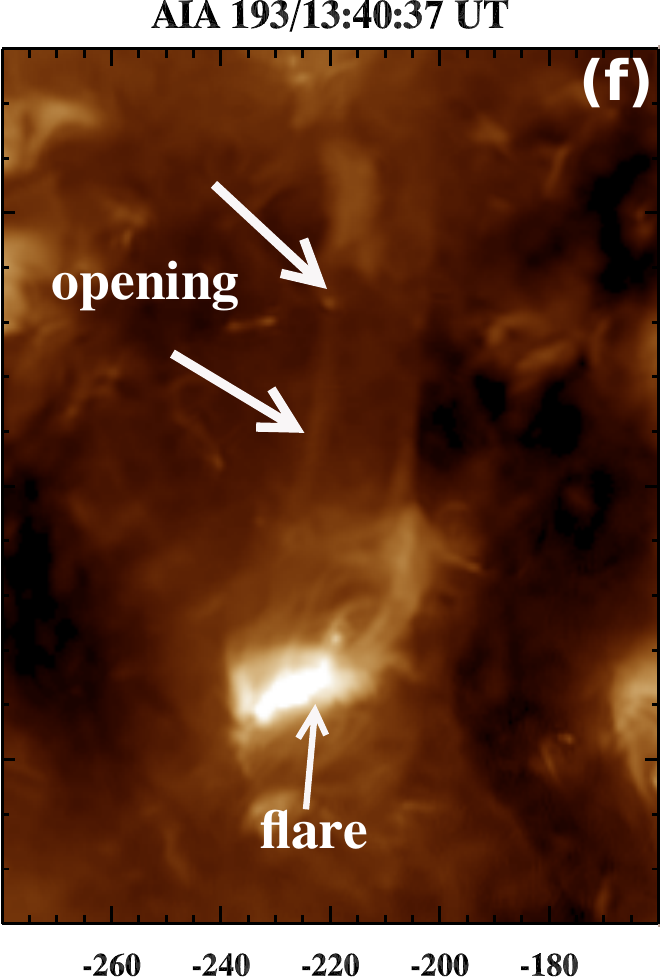}
\includegraphics[width=4.1cm]{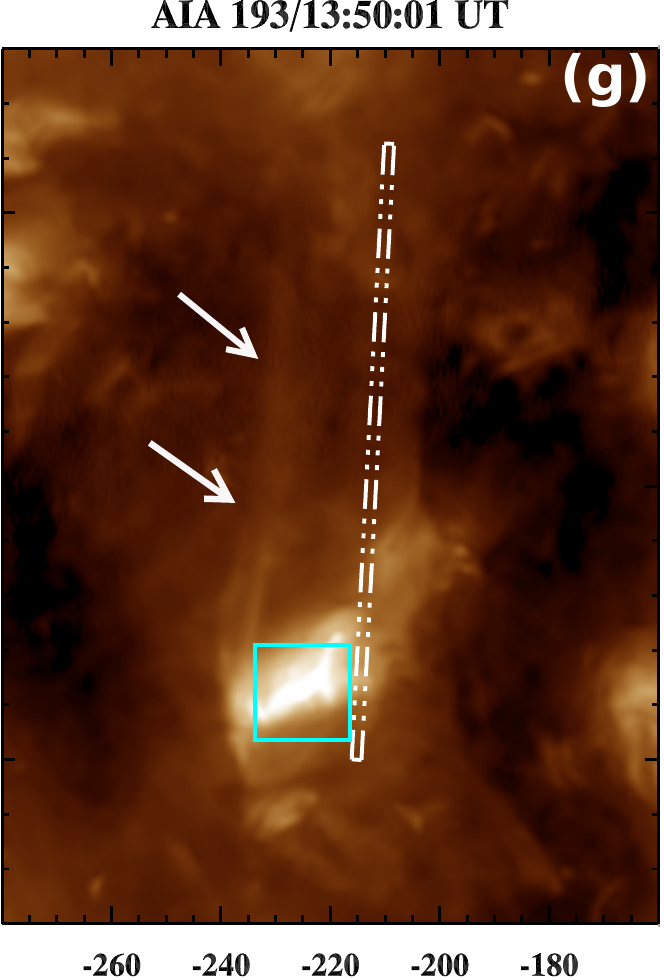}
\includegraphics[width=4.9cm]{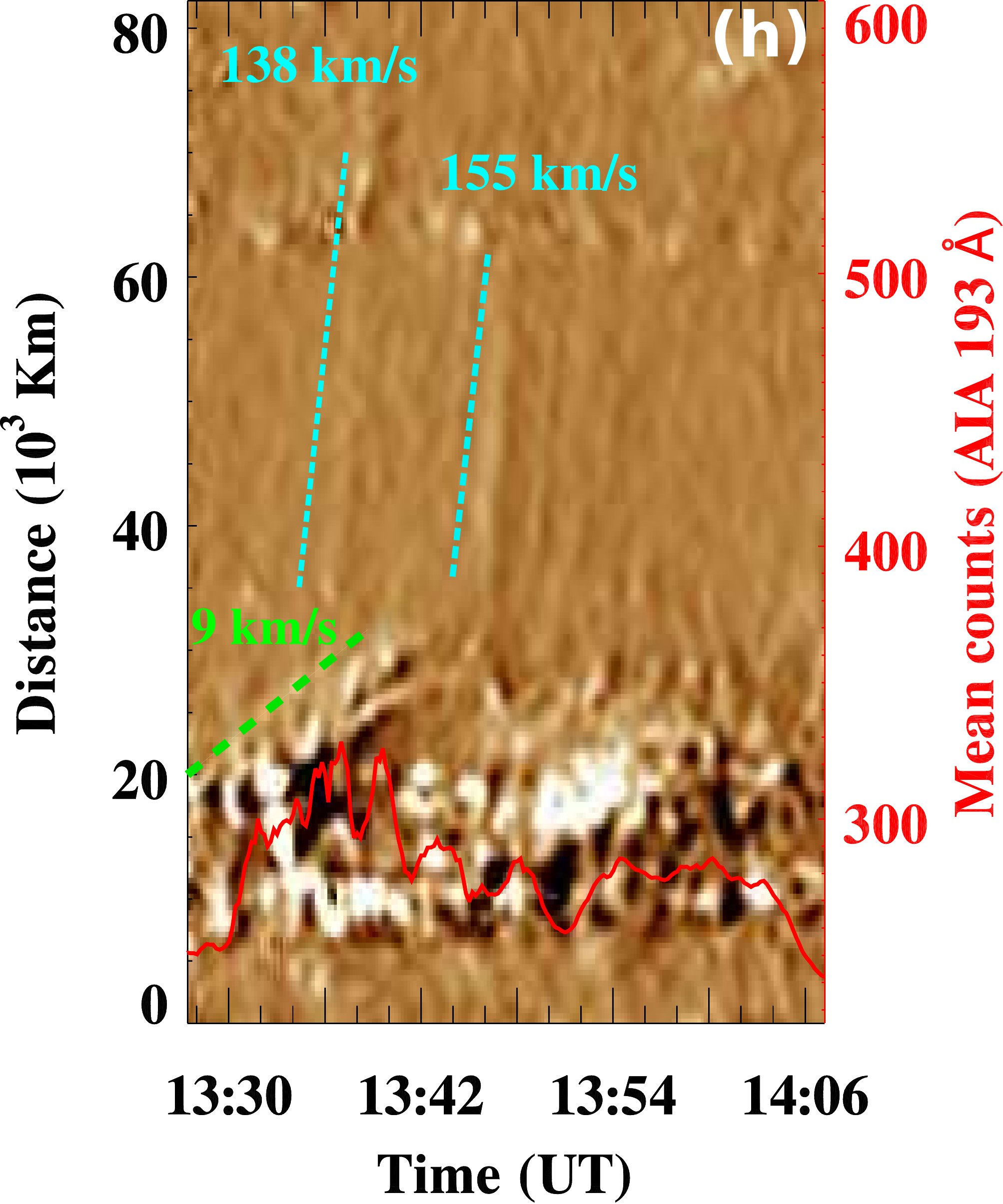}
\includegraphics[width=18cm]{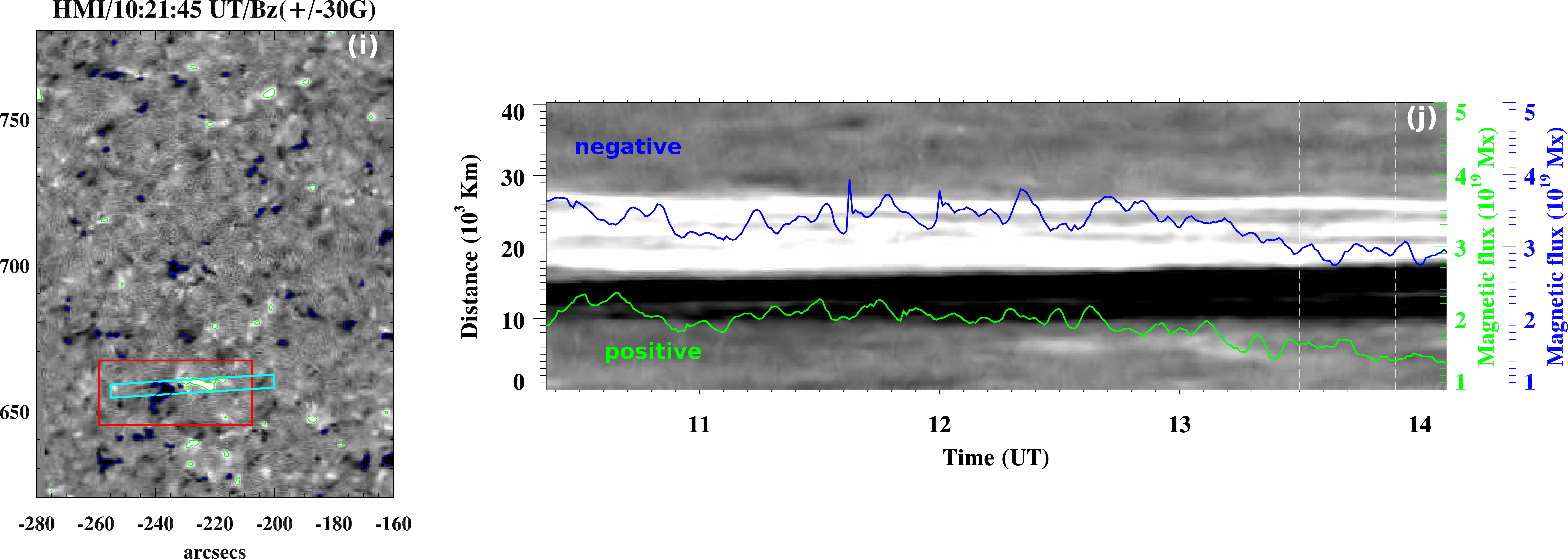}

}
\caption{(a-c) AIA 304, 171, and 193 \AA~ images showing a jet without an associated mini-filament eruption (\#6 in Table 1).  HMI magnetogram contours ($\pm$30 G) of positive (blue) and negative (green) polarities are superposed on (a). The white arrow indicates the location of the jet onset and opening of the rising structure. (d) Potential-field extrapolation of the jet source region, showing selected closed (yellow) and open (red) field lines.  (e) AIA 193 \AA~ base-difference (13:37:49 UT - 13:26:37 UT) image with key features as marked. (f,g) AIA 193 \AA~ intensity images showing the jet origin, newly opened flux, and mini-flare arcade. (h) TD intensity plot extracted from AIA 193 \AA~ images along the slit (white dot-dashed line) shown in (g).  The red curve represents the averaged AIA 193 \AA~ counts from within the cyan box in (g). (i) HMI magnetogram with $\pm$30 G contours prior to the jet onset.  (j) TD flux plot from a 4-hr series of HMI magnetograms, taken from the narrow cyan box in (i) and averaged in the north-south direction. Blue and green curves are the negative and positive fluxes (absolute value) inside the red box in panel (i). Vertical white dashed lines mark the beginning and end of the eruption phase (An animation of this figure is available).   
} 
\label{nof}
\end{figure*}

\subsection{Jets with mini-filament eruptions (Jets \#7 and \#17-19)}

Figure \ref{j1} and accompanying movies demonstrate a jet associated with a mini-filament eruption: Jet \#7 at 14:29:49 UT on 2014 January 8.  HMI magnetogram contours ($\pm$30 G) over the AIA 304 \AA~ image of the source region (Figure \ref{j1}(a)) reveal a central minority-polarity region (positive, green contour) surrounded by majority-polarity concentrations (negative, blue contour). A dark mini-filament lies along the polarity inversion line (PIL) (white arrows in Figure \ref{j1}(a-c)) where the strongest polarities were located (Figure \ref{j1}(a)).  A Potential-field extrapolation of the source region before the jet onset reveals a classic fan-spine topology (Figure \ref{j1}(d)). The selected fan loops and cusp structures near the null (red lines) correspond well to bright features observed in AIA 193 \AA~ (Figure \ref{j1}(c)); the yellow lines below the fan represent field lines overlying the mini-filament.

The TD plot of the averaged AIA 193 \AA~ intensity (Figure 3(j)) along the slit shown in panel (e) provides an overview of the entire event during 14:00-15:28 UT. The overplotted red curve represents flaring activity, as measured by the averaged counts extracted between the horizontal yellow lines on the TD plot.  Before the mini-filament began to rise, the bright point became activated as follows. Between $\approx$14:14 and 14:18 UT, a dark arch (A) and bright loops overlying the northern end of the mini-filament (F) rose slowly ($\approx$13 \kms), while small brightenings appeared underneath (Figure \ref{j1}(e) and (j)).  Diffuse quasiperiodic jetting began at $\approx$14:19 UT, when the overlying structures reached the fan. The driving force responsible for the expansion is unclear from the observations, although the presence of small EUV emission sites beneath the rising loops implies that reconnection was involved. Because the overlying dark arches and bright loops were oriented roughly perpendicular to the PIL, these structures were weakly sheared and contained little free energy. We infer that the expansion of the overlying flux stressed the null at the cusp, forming a breakout current sheet there.  When the rising flux encountered the breakout sheet, reconnection with the adjacent open flux expelled weak jets repeatedly from the vicinity of the spine (see Figure \ref{j1}(j) and accompanying AIA animation).  

From 14:24-14:46 UT, the mini-filament rose slowly ($\approx$10 \kms) accompanied by localized brightenings below (Figure \ref{j1}(e)-(f)) and a leftward drift of the spire. The AIA 193 \AA~ base-difference image at 14:36:49 UT (Figure \ref{j1}(g)) shows this increased activity more clearly than the undifferenced intensity images. At this time, bright fan loops and the first fast jet coincided with a dimming region at the site of the overlying loops and downflows along the surrounding fan loops (visible in the accompanying AIA 193 \AA~ animation). We interpret the downflows and heating in the bright fan loops as consequences of fast breakout reconnection. The dimming region can be attributed to the density depletion resulting from rapid expansion and opening of the upper flux rope surrounding the mini-filament \citep[e.g.,][]{innes2010}.  A series of fast jets (projected speeds of $\approx$220$\pm$15 \kms; see Figure \ref{j1}(j)) were observed during 14:44-15:00 UT. The mini-filament reached the breakout sheet at $\approx$14:38 UT, while the bright postflare arcade, the standard signature of reconnection below the filament, first appeared at 14:40 UT and persisted until $\approx$15:20 UT (Figure \ref{j1}(h)). Therefore strong flare reconnection began significantly after the onset of fast breakout reconnection, in contrast to some other events (e.g., Jet \#14).

The evolution of the photospheric magnetic field starting about 3 hours before the jet onset (11:30-15:20 UT) is depicted by the TD magnetic flux plot of the negative and positive magnetic fluxes inside the red box in Figure \ref{j1}(i), averaged in the north-south direction (in Figure \ref{j1}(k)); the superposed blue and green curves track the absolute values of the negative and positive fluxes within the $\pm$30 G contours inside the red box in Figure \ref{j1}(i).  The negative and positive fluxes began to decrease about 2 hours before the fast jets  ($\approx$12:45 UT), reaching $\approx$67$\%$ and $\approx$50$\%$, respectively, of their initial values by the onset of eruption. There was no significant change in either flux during the jets.  Close inspection of the HMI flux TD plot reveals the disappearance of positive and negative flux patches at different locations (marked by 1 and 2) at $\approx$13:30 UT. The negative patch (1) moves toward the PIL and appears to cancel with opposite polarity, without producing any jet at that time. In contrast, the positive polarity (2) is isolated and most likely submerges or diffuses rather than cancels. Therefore, the separate disappearances of 1 and 2 jointly contribute to the steady decrease in positive/negative fluxes until $\sim$13:00 UT. Later on, shrinkage and submergence of individual polarities contribute to the overall decrease in magnetic flux. Clearly flux cancellation does not trigger the jet, in this case.   

We also observed more complex jets (e.g., Jets \#17-19) associated with adjoining sources and multiple filament eruptions. Figure \ref{twin}(a-c) displays AIA 304, 171, and 193 \AA~ images at 08:21:01 UT showing two source regions (neighboring bright points) in the ECH on 2014 January 10. P1 and P2 (Figure \ref{twin}(a)) are the minority polarities (positive) surrounded by the CH background field. The Potential-field extrapolation of the HMI magnetogram 15 min before the first eruption (Figure \ref{twin}(d)) reveals the classical fan-spine topologies for both bright points. The yellow (closed) field lines connect P1 and P2 to the surrounding opposite polarities, while the red and green (open) field lines outline the two fans. Inspection of the HMI movie shows that continuous footpoint motions produced an inverse S-shaped structure along the PIL containing mini-filament F1 (Figure \ref{twin}(b,c)).  Figure \ref{twin}(k) displays the TD magnetic flux plot along the rectangular yellow slit marked in panel (i); the superposed blue and green curves track the absolute values of the negative and positive fluxes within the $\pm$30 G contours inside the red box in Figure \ref{twin}(i). There was no significant flux emergence or cancellation 1 hour before or during any jet onset as indicated by the roughly steady amount of positive flux (green) after 7:30 UT, although P2 moves continuously toward the nearest concentration of opposite polarity. The accompanying HMI movie shows ongoing diffusion of P2 at the PIL prior to and during the third jet, after which P2 and its associated bright point disappeared.  

The jets were associated with the eruptions of mini-filaments F1 (an inverse S-shaped filament), F2, and F3. Two eruptions occurred almost simultaneously in the neighboring source regions, producing two jets; one hour later, a third jet associated with another mini-filament eruption occurred in the smaller, southern bright point. Figure \ref{twin}(e) shows the eruption of F1 (Jet \#17) at 08:23:13 UT in AIA 193 \AA~ producing a quasi-circular ribbon at the base of the fan and a diffuse jet (see accompanying animation for details). The projected jet speed was $\approx$50 \kms, but the actual jet speed was probably much higher due to its alignment with the line of sight. As shown in the TD intensity plot (Figure \ref{twin}(j)) along the curved slit outlined in Figure \ref{twin}(e), the first jet was preceded by the activation and slow rise (for about 1 hour) of F1 with a speed of $\approx$3 \kms. We did not observe any brightening below F1 during this phase. 

Shortly after F1 erupted, F2 rose slowly ($\approx$13 \kms) within the neighboring southern bright point. A cusp-shaped bright structure, circular ribbon at the fan base, quasiperiodic narrow jets (Jet \#18: $v\approx$108$\pm$25 \kms), and diffuse outflows formed over the next 20-30~min (Figure \ref{twin}(f) and (g)). Finally, F3 rose slowly ($\approx$5 \kms) and erupted at $\approx$09:32 UT (Jet \#19) from the same PIL as F2 (Figure \ref{twin}(h)), producing a narrow jet ($\approx$86$\pm$26 \kms) in a manner similar to the prior eruptions. We interpret this activity as the consequence of breakout reconnection between the flux ropes supporting the filaments and the external open field, as we demonstrated earlier for Jet \#14 \citep{kumar2018}. The ejections of F2 and F3 show that multiple mini-filament eruptions from the same source region can drive sequential jets, most likely associated with the eruption of different segments of the same filament channel. It is also possible that the F1 eruption triggered that of F2, based on their close spatial and temporal proximities, as in sympathetic flares/CMEs. 

\subsection{Jets without mini-filament eruptions (Jet \#6)}

Figure \ref{nof} and the accompanying animation show a jet during 13:30-14:00 UT on 2014 January 8 that lacks visible signs of a filament in the AIA channels (Figure \ref{nof}(a-c)). The Potential-field extrapolation of the jet source before the eruption (Figure \ref{nof}(d)) reveals an asymmetric fan-spine topology, where most of the closed (yellow) and open (red) flux was concentrated on the left side of the fan. As shown by the TD intensity plot (Figure \ref{nof}(h)) extracted from a slice (white dot-dashed line in Figure \ref{nof}(g)) in the AIA 193 \AA~ running-difference images, the closed structures on the left side expanded slowly ($\approx$9 \kms) during 13:28-13:40 UT. The AIA 193 \AA~ base difference image at 13:37:49 UT (Figure \ref{nof}(e)) reveals brightening close to the magnetic null, which we attribute to compression and distortion of the null by the expanding closed structures, creating the breakout current sheet. Repeated collimated jets were produced by this region from $\approx$13:36 UT onward, accompanied by successive left-to-right deflections of the spire that we interpret as a signature of breakout reconnection. Coronal dimmings (Figure \ref{nof}(e)), a small flare arcade (Figure \ref{nof}(f,g)), and quasiperiodic flows beneath the rising structure (Figure \ref{nof}(h)) also were detected during this phase.

At the same time as the jets were expelled with speeds of $\approx$138 and 155 \kms (Figure \ref{nof}(h)), the interior intensity of the closed bright structures also varied, indicating that individual episodes of breakout reconnection coincided with episodes of fast flare reconnection below the rising structure.  The TD plot (Figure \ref{nof}(j)) of the fluxes along a narrow slit (cyan) in the HMI magnetogram (Figure \ref{nof}(i)) during 10:28-14:06 UT, and the positive and negative fluxes measured within the red box on Figure \ref{nof}(i), exhibit no significant flux emergence or cancellation during the 3.5 hours before the eruption began. No magnetic changes were seen during the subsequent jet phase (between the white vertical dashed lines). Although the opposing polarities in the red box converged, negligible flux was cancelled.


\section{DISCUSSION AND CONCLUSIONS}\label{conclusions}
All 27 ECH jets on 2013 June 27-28 and 2014 January 8-10 that we analyzed occurred in embedded-bipole fan-spine topologies, which we have investigated intensively through observations and numerical simulations as the source of reconnection-driven jets \citep{pariat2009,pariat2010,pariat2015,pariat2016,wyper2016a,wyper2016b,karpen2017,kumar2018}. If the majority-polarity magnetic flux surrounding the central minority polarity were symmetrically distributed, a circular filament channel would form along the circular PIL. Most of the analyzed events exhibit filament formation only on one side of the central minority polarity, however, because the surrounding opposite polarities are asymmetrically distributed. Our simulations \citep[e.g.,][]{wyper2018} and observations consistently demonstrate that the erupting section of a filament channel preferentially forms at the PIL between the strongest polarities, which is generally the inner PIL of the emerged flux.

Most of our selected jets ($\approx$67\%) were associated with mini-filament eruptions, while the remaining third do not contain mini-filaments but manifest clear symptoms of flare reconnection during the event. The jets associated with filament eruptions exhibit: (1) a slowly rising filament ($\approx$10 \kms), characteristic of a flux rope forming and rising above the PIL; (2) quasiperiodic, diffuse, straight jets from the vicinity of the spine and remote brightenings aligned with the base of the fan, characteristic of slow breakout reconnection between the closed flux above the filament and the external open field; (3) explosive reconnection at a flare current sheet below the filament, producing flare arcades and accelerating the rising flux rope; and (4) explosive breakout reconnection of the flux rope with open CH flux, generating helical jets with typical speeds $\approx$100-400 \kms that contain both hot and cool plasma. Two well-resolved events in our sample reveal plasmoids in the flare current sheet below the rising filament/flux rope. All of these features are consistent with the breakout model of solar jets \citep{wyper2017,wyper2018}. 

The jets without mini-filament eruptions exhibit similar observable signatures as those with mini-filament eruptions:  slowly rising structures inside the fan, a flare arcade, remote brightenings, plasmoids, and coronal dimming regions. However, these filament-free jets are not as violent as those associated with mini-filament eruptions, presumably because the amounts of magnetic shear and associated free energy are lower than in filament-containing events. The general lack of helical motion also is consistent with a smaller amount of shear/twist driving these events. This observational evidence strongly suggests that jets without mini-filament eruptions simply contain a filament channel without cool material, and the channel magnetic flux is partially converted by flare reconnection to a flux rope as in the filament-eruption cases. 

{\it Therefore, all ECH jets (with or without filaments) in our sample are breakout jets exhibiting common observational features.} These results have important implications for the buildup and release of energy in solar eruptions on all scales. Our investigation strongly supports the \citet{sterling2015} discovery that mini-filament (to be precise, mini-filament channel) eruptions drive coronal jets. Even in these small structures, the shear and magnetic free energy become concentrated at the portion of the PIL between the strongest flux concentrations. In addition, we have found no evidence of jets initiated by the resistive-kink mechanism \citep[e.g.,][]{pariat2009} or another global ideal instability of the closed-field system. 

In our study, only 6 out of 27 jets showed even small amounts of flux cancellation (or diffusion) during the 2-3 hours before and during the jet. The remaining events manifested no measurable flux emergence or cancellation associated with the eruption. Flux emergence clearly is a prerequisite for creating the jet source regions. However, the newly emerged bipoles do not erupt immediately, which indicates that they emerge with insufficient free energy to enable eruption. Therefore, we conclude that shearing and/or rotational photospheric motions are the most likely sources of the energy buildup that forms the filament channel and is released through eruption \citep[e.g.,][]{wyper2018a}, as in the helicity condensation model \citep{antiochos2013,knizhnik2017,dahlin2018}. Some of the larger analyzed events clearly exhibit rotational and/or shearing displacements between the minority polarity and its surroundings (e.g., see animations of Figs. \ref{bp} and \ref{twin}). However, discerning such motions in the small, magnetically weak jet sources with only line-of-sight magnetograms is difficult with current instrumentation, particularly for the smaller, poorly resolved jets without mini-filaments. Further work is required to establish whether large-scale rotational motions or the helicity-condensation mechanism can generate the filament channel and the required free energy within the observed intervals between emergence and eruption, particularly for recurrent jets from the same source. 
 
As established by previous research and the present investigation, the general scenario for the coronal-hole jets in our study is as follows. Embedded bipoles emerge in coronal holes, generating a fan-spine topology by connecting the central minority-polarity region with surrounding opposite-polarity field and forming coronal bright points \citep[e.g.,][]{golub1980}. Footpoint motions beneath the fan build up free magnetic energy at the PIL, creating mini-filament channels. The slow rise of the mini-filament forms the flare current sheet and allows the flare reconnection to form a growing flux rope surrounding the cool filament plasma. The slow rise also is enabled by breakout reconnection at the deformed null, which erodes the overlying strapping field.  In principle, flux cancellation at the PIL could contribute to the buildup of the flux rope \citep[e.g.,][]{Kumar2015,kumar2017}.  Well-calibrated observations and focused numerical studies are needed to understand how flux cancellation works on the Sun, however, and its possible contribution to the accumulation and release of free energy in impulsive eruptions. As the pre-eruptive phase proceeds, the unsheared flux above the flux rope reconnects with external open flux through the breakout current sheet, producing diffuse quasiperiodic jets.  When the flare reconnection transitions to a faster rate, the flux-rope rise speed increases by an order of magnitude. After a delay dependent on the height of the breakout sheet and the speed of the rising flux rope, fast breakout reconnection and the expulsion of the Alfv\'enic jet are instigated when the twisted flux contacts the breakout current sheet.  Ideal instability appears to play no role in the eruption; all jets are released by fast reconnection through the null deformed into a current sheet, coupled with fast reconnection in the flare current sheet. As an example of similar behavior on a larger scale, we reported an active-region event in which the S-shaped flux rope erupted $\approx$3 hours after its formation and produced helical jets associated with the appearance of a quasicircular ribbon, in a fan-spine topology \citep{Kumar2015}. Many bright points erupt repeatedly, enhancing their contributions of mass and energy to the corona and solar wind. Finally, the bright point disappears in $\approx$1-6 days (or more) depending on the magnetic field strength, as the minority polarity submerges, diffuses, or cancels. We conclude from our statistical investigation that the breakout mechanism explains most, if not all, ECH jets, and that neither flux cancellation nor emergence plays a leading role in triggering these ubiquitous eruptions. 
\\



\begin{longrotatetable}
\begin{longtable}{c c c c c c c c c c}

\caption{Characteristics of the selected equatorial coronal-hole jets$\dagger$} \\
\hline \\
\label{tab1}

Jet \# &Date  &Eruption &Description &Flare arcade &Filament? &Flux cancellation? &Time interval &Movies  \\
& &Start-End$^{a}$ & & appearance & &   &  between BP   &1=AIA \\
& &Time (UT) & & (UT) & &  & appearance and  &304+171 \\
& && &  & &  &first jet  &+193(12-s) \\
& &Location   & &  & &  & (hrs/days) &2=AIA 304 \\
& & X,Y (arcsec) & &  & &  &  &+HMI (45-s) \\
\hline
\\
   
1 &2014  & 01:00-01:40   & Weak flare arcade at right side of fan; & 01:25  & No  & No &19h &\href{https://zenodo.org/api/files/77d39b7b-c26c-47ec-a5e2-466ab090394d/jet01A.mp4}{1},\href{https://zenodo.org/api/files/77d39b7b-c26c-47ec-a5e2-466ab090394d/jet01B.mp4}{2}    \\    
     &Jan. 8  &(-292,697)   & spire drifts toward left side of fan.   &  &  &  &&   \\
 &  &   &   &  &  &&&     \\

2 &  &  02:45-03:25   & Plasmoid formation and propagation along the jet spire; &03:08   & No     & No &19h &\href{https://zenodo.org/api/files/77d39b7b-c26c-47ec-a5e2-466ab090394d/jet02A.mp4}{1},\href{https://zenodo.org/api/files/77d39b7b-c26c-47ec-a5e2-466ab090394d/jet02B.mp4}{2} \\
&  &(-281, 697)  & flare arcade beneath left side of  fan;    &  &  & &&    \\
&  &   &spire drift toward right side of fan. &  &   &  &&   \\
 &  &   &   &  &  &&&     \\

3 &  &  04:00-04:25   & Initial brightening close to right end of mini-filament; & 04:12    & Yes     & Disappearance of  &19h   & \href{https://zenodo.org/api/files/77d39b7b-c26c-47ec-a5e2-466ab090394d/jet03A.mp4}{1}, \href{https://zenodo.org/api/files/77d39b7b-c26c-47ec-a5e2-466ab090394d/jet03B.mp4}{2} \\
 &  & (-276, 697)   & disconnection of the right leg of mini-filament; &  &  & tiny negative flux &    &   \\
 &  &   & counterclockwise rotation of mini-filament; &  &   & during the jet. &&    \\
&   &  & quasicircular ribbon. &  &   &  &&   \\
 &  &   &  &  &   &    && \\
 
4 &  &  07:20-07:40   & Narrow diffuse jet early;  &07:45   & Yes     & No & $\star$  & \href{https://zenodo.org/api/files/77d39b7b-c26c-47ec-a5e2-466ab090394d/jet04A.mp4}{1}, \href{https://zenodo.org/api/files/77d39b7b-c26c-47ec-a5e2-466ab090394d/jet04B.mp4}{2}\\
&  & (-205, 610)   & slow rise and eruption of S-shaped structure+filament; &  &   & &&    \\
 &  &   & flare arcade beneath right side of  fan;   &  &   & &&    \\
 &  &   & circular ribbon formation; &  &   &  &&   \\
 &  &   & jet spire drift toward left side of fan. &  &   &   &&  \\
 &   &  &  &  &   &   &&  \\

5 &  &  08:00-09:25   & Expansion and eruption of rising  & 08:25   & No     & No  &4h & \href{https://zenodo.org/api/files/77d39b7b-c26c-47ec-a5e2-466ab090394d/jet05A.mp4}{1}, \href{https://zenodo.org/api/files/77d39b7b-c26c-47ec-a5e2-466ab090394d/jet05B.mp4}{2}\\
&   & (-238, 663)  &structure beneath left side of the fan;  &  &   &  &&   \\
  &   &  & straight narrow jet.   &  &   &   &&  \\
  &   &  &  &  &   &  &&   \\

6 &  &  13:30-14:00   & Flare arcade on left side of fan; &13:45 & No     & No  & 4h &\href{https://zenodo.org/api/files/77d39b7b-c26c-47ec-a5e2-466ab090394d/jet06A.mp4}{1}, \href{https://zenodo.org/api/files/77d39b7b-c26c-47ec-a5e2-466ab090394d/jet06B.mp4}{2} \\
 & & (-208, 663)  & quasiperiodic outflows below rising structure;  &    &  & &&      \\
 & &  & successive opening of the fan;  &    &  &    &&   \\
 & &  &   straight narrow jet.  &  &  &  && \\
&   &  &  &  &   &   &&  \\

7 &  &  14:10-15:25   & Mini-filament+overlying arcade eruption; &14:40 & Yes     & Tiny   &20h   & \href{https://zenodo.org/api/files/77d39b7b-c26c-47ec-a5e2-466ab090394d/jet07A.mp4}{1},\href{https://zenodo.org/api/files/77d39b7b-c26c-47ec-a5e2-466ab090394d/jet07B.mp4}{2}\\
 &   & (-118, 683)  & diffuse jets driven by rising arcade   &  & &   cancellation    &&   \\
 &   &  &above mini-filament;  &  &   &    &&  \\
&   &  &flare arcade beneath right side of fan.   &  &   &   &&  \\
&   &  &  &  &   &   &&  \\

8 &  &  14:40-15:30   & Diffuse jet onset prior to mini-filament eruption; &15:04 & Yes    & No &$\star$ &\href{https://zenodo.org/api/files/77d39b7b-c26c-47ec-a5e2-466ab090394d/jet08A.mp4}{1}, \href{https://zenodo.org/api/files/77d39b7b-c26c-47ec-a5e2-466ab090394d/jet08B.mp4}{2} \\
 &   & (-114, 552)  &Filament+overlying arcade eruption;  &  &   &   &&  \\
&  &     & disconnection of one leg of mini-filament; &      &  &  &&  \\
&   &  & double jet spire; &  &   &   &&  \\
 &  &     & counterclockwise rotation of mini-filament inside fan;  &      &  & &&   \\
 &   &  & flare arcade beneath the right side of fan. &   &    &  &&   \\
 &   &  & &  &   &   &&  \\

9 &  &  16:15-16:50   & Mini-filament+overlying arcade eruption; &16:24 & Yes     &  Repeated appearance &15h   &\href{https://zenodo.org/api/files/77d39b7b-c26c-47ec-a5e2-466ab090394d/jet09A.mp4}{1}, \href{https://zenodo.org/api/files/77d39b7b-c26c-47ec-a5e2-466ab090394d/jet09B.mp4}{2}  \\
&   & (-85, 744)  & flare arcade beneath left side of fan; & &   & and disappearance of  &   &   \\
&   &  &spire drifts to right side of fan. &  &   &negative element.   &&  \\
&   &  &  &  &   &   &&  \\

10 &  &  20:53-21:15   & Initial brightening below mini-filament; &21:05 & Yes     &  No &$\star$ & \href{https://zenodo.org/api/files/77d39b7b-c26c-47ec-a5e2-466ab090394d/jet10A.mp4}{1}, \href{https://zenodo.org/api/files/77d39b7b-c26c-47ec-a5e2-466ab090394d/jet10B.mp4}{2}  \\
&   & (-91, 699)  & flare arcade beneath left side of fan. &   &    & &&     \\
 &   &  &  &  &   &    && \\

11 &  &  22:00-22:55   & Initial brightening below mini-filament; &22:30 & Yes     & No &20h  & \href{https://zenodo.org/api/files/77d39b7b-c26c-47ec-a5e2-466ab090394d/jet11B.mp4}{1}, \href{https://zenodo.org/api/files/77d39b7b-c26c-47ec-a5e2-466ab090394d/jet11B.mp4}{2} \\
 &  & (-64, 684)     & spire drifts to left side of fan; &     &  &  &&  \\
&   &  & flare arcade beneath right side of fan. &   &    &   &&  \\
 &   &  &  &  &   &  &&   \\

12 & 2014 & 04:55-05:12   & Flare arcade beneath right side of fan; &05:01 & Yes     & No &3h   &  \href{https://zenodo.org/api/files/77d39b7b-c26c-47ec-a5e2-466ab090394d/jet12B.mp4}{1},\href{https://zenodo.org/api/files/77d39b7b-c26c-47ec-a5e2-466ab090394d/jet12B.mp4}{2}\\ 
     &Jan. 9  &(-34, 522)   & reconnection above and below the filament;  & &  & &&    \\
 &  &    &      untwisting mini-filament;   &  &  &   &&  \\
 &  &   & double jet spire.   &  &  &  &&   \\
&   &  &  &  &   &   &&  \\

13 &  & 08:42-09:05   & Flare arcade beneath left side of  fan; &08:47 & Yes     & No &17h & \href{https://zenodo.org/api/files/77d39b7b-c26c-47ec-a5e2-466ab090394d/jet13A.mp4}{1},\href{https://zenodo.org/api/files/77d39b7b-c26c-47ec-a5e2-466ab090394d/jet13B.mp4}{2} \\  
   &  & (60, 581)   & Breaking of mini-filament at fan apex;  &  &  &  &&   \\
 &  &   & quasiperiodic ejections;  &  &  &  &&   \\
&   &  &quasicircular ribbon;  &  &   &   &&  \\
&   &  &upflows and downflows in flare current sheet. &  &   &   &&  \\
&   &  &  &  &   &   &&  \\

14 &  & 16:45-17:23   & Slow rise of mini-filament; multiple plasmoids &17:18  & Yes     & No  & 5d &\href{http://adsabs.harvard.edu/abs/2018ApJ...854..155K}{Kumar et al. 2018} \\  
     &  &(150, 405)   & in flare current sheet; flux rope formation;  &  &  & &&    \\
 &   &  &  helical jet. &  &   &   &&  \\
&   &  &  &  &   &  &&   \\

15 &2014  & 07:57-08:40   & Flare arcade beneath left side of fan; &08:20 & No     & No &12h &\href{https://zenodo.org/api/files/77d39b7b-c26c-47ec-a5e2-466ab090394d/jet15A.mp4}{1},\href{https://zenodo.org/api/files/77d39b7b-c26c-47ec-a5e2-466ab090394d/jet15B.mp4}{2}  \\
      &Jan. 10~ & (-162, 718)   & quasiperiodic downflows and upflows  &  &  &   &&  \\
&   &  & during jet. &  &   &   &&  \\
&   &  &  &  &   &   && \\

16 &  & 08:25-08:50   & Filament slow rise associated with  &08:37 & Yes     & No &4.5h  &\href{https://zenodo.org/api/files/77d39b7b-c26c-47ec-a5e2-466ab090394d/jet16A.mp4}{1},\href{https://zenodo.org/api/files/77d39b7b-c26c-47ec-a5e2-466ab090394d/jet16B.mp4}{2}    \\
      &  & (246, 419)   & brightening inside the fan;  &  &  &  &&   \\
&   &  & counterclockwise rotation of jet.  &  &   &   &&  \\
&   &  &  &  &   &   &&  \\

17 &  & 08:05-08:32   & North: Slow rise of mini-filament; &08:28 & Yes     & No &22h  & \href{https://zenodo.org/api/files/77d39b7b-c26c-47ec-a5e2-466ab090394d/jet17-19A.mp4}{1},\href{https://zenodo.org/api/files/77d39b7b-c26c-47ec-a5e2-466ab090394d/jet17-19B.mp4}{2}   \\
 &   & (186, 260)  &  circular flare ribbon;  &  &   &  &&   \\
&   &  & counterclockwise rotation of jet.  &  &   &   &&  \\
&   &  &  &  &   &  &&   \\

18 &  & 08:25-08:56   & South: Mini-filament eruption &08:35 & Yes     & No &5h  &\href{https://zenodo.org/api/files/77d39b7b-c26c-47ec-a5e2-466ab090394d/jet17-19A.mp4}{1},\href{https://zenodo.org/api/files/77d39b7b-c26c-47ec-a5e2-466ab090394d/jet17-19B.mp4}{2}   \\
      &  & (198, 242)   & with narrow jet;  &  &  &  &&   \\
&   &  & circular flare ribbon;  &  &   &  &&   \\
&   &  & counterclockwise rotation of jet.  &  &   &   &&  \\
&   &  &  &  &   &  &&   \\

19 &  & 09:30-09:50   & South: Mini-filament &09:37 & Yes     & No &5h &\href{https://zenodo.org/api/files/77d39b7b-c26c-47ec-a5e2-466ab090394d/jet19C.mp4}{193}   \\
      &  & (198, 242)   & eruption 1 hour after Jet \#18  &  &  &  &&   \\
     &  &   & originating from same PIL;  &  &  &  &&   \\
&   &  & circular flare ribbon.  &  &   &   &&  \\
&   &  &  &  &   &  &&   \\

20 &  & 17:44-18:06   & North: Weak flare arcade beneath left side of  fan; &18:00 & No     & No &8h &\href{https://zenodo.org/api/files/77d39b7b-c26c-47ec-a5e2-466ab090394d/jet20A.mp4}{1},\href{https://zenodo.org/api/files/77d39b7b-c26c-47ec-a5e2-466ab090394d/jet20B.mp4}{2}   \\
      &  & (360, 429)   &  quasiperiodic outflow along  jet spire.  &  &  & &&    \\
&   &  &  &  &   &  &&   \\

21 &  & 17:51-18:12   & South: Flare arcade beneath left side of fan; &17:55 & No     & No &8h  &\href{https://zenodo.org/api/files/77d39b7b-c26c-47ec-a5e2-466ab090394d/jet21A.mp4}{1},\href{https://zenodo.org/api/files/77d39b7b-c26c-47ec-a5e2-466ab090394d/jet21B.mp4}{2}   \\
      &  & (342, 393)  & multiple tiny plasmoids along the jet spire.  &  &  & &&    \\
&   &  &  &  &   &  &&   \\

22 &  & 18:00-18:20   & North: flare arcade beneath north side of fan;  &18:10 & Yes     & Yes &16h  &\href{https://zenodo.org/api/files/77d39b7b-c26c-47ec-a5e2-466ab090394d/jet22A.mp4}{1},\href{https://zenodo.org/api/files/77d39b7b-c26c-47ec-a5e2-466ab090394d/jet22B.mp4}{2}   \\
      &  & (335, 322)   & minor flux cancellation  &  &  &  &&   \\
&   &  & at initial brightening site below mini-filament; &  &   &  &&   \\
&   &  & clockwise rotation of jet.  &  &   &  &&   \\
&   &  &  &  &   &  &&   \\

23 &  & 18:30-18:55   & South: flare arcade beneath south side of fan;   &18:40 & Yes     & Yes &18h  &\href{https://zenodo.org/api/files/77d39b7b-c26c-47ec-a5e2-466ab090394d/jet23A.mp4}{1},\href{https://zenodo.org/api/files/77d39b7b-c26c-47ec-a5e2-466ab090394d/jet23B.mp4}{2}   \\
      &  & (367, 181)   & minor flux cancellation  &  &  &  &&   \\
&   &  &  below the filament; &  &   &  &&   \\
&   &  & clockwise rotation of jet.  &  &   &  &&   \\
&   &  &  &  &   &  &&   \\

24 &  & 04:32-04:50   & Flare arcade beneath right side of fan; &04:45  & No     & No &12h  &\href{https://zenodo.org/api/files/77d39b7b-c26c-47ec-a5e2-466ab090394d/jet24A.mp4}{1},\href{https://zenodo.org/api/files/77d39b7b-c26c-47ec-a5e2-466ab090394d/jet24B.mp4}{2}    \\
      &  & (-185, 717)   & weak remote brightening.  &  &  &  &&   \\
&   &  &  &  &   &   &&  \\

25 &  & 01:40-02:15   & Jet from right side of fan;  &01:45 & No     & No &5d &\href{https://zenodo.org/api/files/77d39b7b-c26c-47ec-a5e2-466ab090394d/jet25A.mp4}{1},\href{https://zenodo.org/api/files/77d39b7b-c26c-47ec-a5e2-466ab090394d/jet25B.mp4}{2}  \\
      &  & (226, 397)   & weak flare arcade;  &  &  &  &&   \\
&   &  &upflows and downflows during jet.   &  &   &   &&  \\
&   &  &  &  &   &   &&  \\

26 &2013   & 00:00-01:50   & Repeated jets every 8-10 min; & Too tiny  & Yes     & Yes & 2 h, & \href{https://zenodo.org/api/files/77d39b7b-c26c-47ec-a5e2-466ab090394d/jet26A.mp4}{1},\href{https://zenodo.org/api/files/77d39b7b-c26c-47ec-a5e2-466ab090394d/jet26B.mp4}{2}\\
 &June 27~  & (240, 124)   & elongated dimming along outer spine; & to detect &  &   &&  \\
&   &  & convergence and cancellation between negative  & the arcade &   &  &&   \\
&   &  & and surrounding positive polarities; & (whole fan &   &   & &  \\
&   &  &only last jet associated with mini-filament eruption;  & brightening&& &     &     \\
&   &  &no jet during rapid negative flux decrease. &     during jets)&   &&&  \\
&   &  & &  &     &&&   \\
&   &  &  &  &     &&&   \\

27 & 2013 & 07:00-07:40   & S-shaped mini-filament eruption with two ribbons; &07:30  & Yes     & Yes &3d & \href{https://zenodo.org/api/files/77d39b7b-c26c-47ec-a5e2-466ab090394d/jet27A.mp4}{1},\href{https://zenodo.org/api/files/77d39b7b-c26c-47ec-a5e2-466ab090394d/jet27B.mp4}{2}\\
   &June 28~   & (486, 192)   & initial brightening below the filament; &  &  &   &&  \\
      &  &   & quasicircular remote ribbon; &  &  &  &&   \\
&   &  & untwisting motion (clockwise) of mini-filament;  &  &   &  &&   \\
&   &  &  dimming regions at ends &  &   &   &&  \\
&   &  & of mini-filament.  &  &   &  &&   \\
&   &  &  &  &   &  &&   \\

\hline
\end{longtable}
\small
\noindent
${}^\star$ The exact time of BP appearance (for 3 events) is unclear due to weak EUV emission and magnetic field.\\
${}^\dagger$ The snapshots of all jets along with supplemental animation files (AIA and HMI movies) are available in the Zenodo repository at: \dataset[DOI]{https://zenodo.org/api/files/77d39b7b-c26c-47ec-a5e2-466ab090394d/list.pdf}.\\ 
${}^a$ The eruption start time and end time of jet is estimated from the AIA 193 \AA~ images.

\end{longrotatetable}

\acknowledgments
We thank N. Raouafi and V. Uritsky for stimulating discussions. SDO is a mission for NASA's Living With a Star (LWS) program. This research was supported by PK's appointment to the NASA Postdoctoral Program at the Goddard Space Flight Center, administered by the Universities Space Research Association through a contract with NASA, and by a grant from the NASA Heliophysics Supporting Research program. PFW was supported through an award of a Royal Astronomical Society Fellowship. JTK, SKA, CRD, and CED were supported by grants from NASA's H-SR, H-LWS, and H-ISFM programs. Magnetic field extrapolations were produced by VAPOR (www.vapor.ucar.edu), a product of the Computational Information Systems Laboratory at the National Center for Atmospheric Research.


\bibliographystyle{aasjournal}
\bibliography{reference.bib}


\end{document}